\documentclass[5p]{elsarticle}

\usepackage{hyperref}
\usepackage{amsmath}
\usepackage{gensymb}
\usepackage{csquotes}

\journal{Astronomy and Computing}

\biboptions{authoryear}

\begin{document}

\begin{frontmatter}
	
	\title{Point Source Detection and False Discovery Rate Control on CMB Maps}
	
	\author[add1,correspondingauthor]{Carr\'on Duque, Javier}
	\ead{javier.carron@uniroma1.it}
	\author[add1,add5]{Buzzelli, Alessandro}
	\ead{alessandro.buzzelli@roma2.infn.it}
	\author[add2]{Fantaye, Yabebal}
	\ead{yabi@aims.ac.za}
	\author[add3]{Marinucci, Domenico}
	\ead{marinucc@mat.uniroma2.it}
	\author[add4]{Schwartzman, Armin}
	\ead{armins@ucsd.edu}
	\author[add1,add5]{Vittorio, Nicola}
	\ead{nicola.vittorio@uniroma2.it}
	
	\address[add1]{Dipartimento di Fisica, Universita' di Roma Tor Vergata, via della Ricerca Scientifica 1, I-00133, Roma, Italy}
	\address[add2]{African Institute for Mathematical Sciences and Department of Mathematics, University of Stellenbosch, South Africa}
	\address[add3]{Dipartimento di Matematica, Universita' di Roma Tor Vergata, via della Ricerca Scientifica 1, I-00133, Roma, Italy}
	\address[add4]{Division of Biostatistics, University of California, San Diego, USA}
	\address[add5]{Sezione INFN Roma 2, via della Ricerca Scientifica 1, I-00133, Roma, Italy}
	
	\cortext[correspondingauthor]{Corresponding author}

\begin{abstract}
	We discuss a new procedure to search for point sources in Cosmic Microwave background maps; in particular, we aim at controlling the so-called False Discovery Rate, which is defined as the expected value of false discoveries among pixels which are labelled as contaminated by point sources. We exploit a procedure called STEM, which is based on the following four steps: 1) needlet filtering of the observed CMB maps, to improve the signal to noise ratio; 2) selection of candidate peaks, i.e., the local maxima of filtered maps; 3) computation of \emph{p-}values for local maxima; 4) implementation of the multiple testing procedure, by means of the so-called Benjamini-Hochberg method. Our procedures are also implemented on the latest release of Planck CMB maps.
\end{abstract}


\end{frontmatter}


\section{Introduction}

In the last decades, observations and characterization of Cosmic Microwave Background (CMB) temperature ani\-so\-tropy established the foundations for the $\Lambda$CDM concordance model (see, e.g., BOOMERanG: \citealt{mactavish2006}; WMAP: \citealt{hinshaw2013}; Planck: \citealt{planckcollaboration2016a}). According to this scenario, the Universe is described by a flat Euclidean geometry, with cosmic structures originated by an almost scale-invariant spectrum of adiabatic Gaussian primordial fluctuations, and the cosmic energy density is in the form of barionic matter ($\sim 5\%$), cold Dark Matter ($\sim 26.5\%$) and Dark Energy ($\sim 68.5\%$).

However, an accurate analysis of the CMB polarization pattern is required in order to break the degeneracy among some parameters and to constrain critical aspects of the Early Universe.
CMB polarization is usually decomposed into a gradient and a curl component, so-called E and B modes, respectively \citep{kamionkowski1997}. E-modes have been widely detected \cite[see, e.g.,][]{kovac2002, planckcollaboration2016b}, while primordial B-modes have escaped observation so far \citep{bicepkpc2015}. The detection of the B-mode polarization represents nowadays the new frontier of observational cosmology, as B-modes would provide an ultimate confirmation to the existence of a stochastic primordial background of gravitational waves, as predicted by inflationary models \citep{lyth1999}.

The primary concern in CMB observations is the presence of foreground contamination, due to both diffuse Galactic emission (e.g. \citealt{planckcollaboration2016c}) and extragalactic (point) sources (e.g. \citealt{planckcollaboration2016d}). In intensity, Galactic emission consists in a mixture of many processes, such as free-free emission, anomalous microwave emission eventually due to non-thermal (spinning) dust and synchrotron emission at low frequency, and thermal dust emission at high frequency. Extragalactic sources, as well, have different astrophysical origin. In particular, the two main classes of sources are radio galaxies and dusty galaxies that are present mainly at low and high frequencies, respectively.

To disentangle the CMB signal from foreground contamination, many approaches have been followed \citep{planckcollaboration2018a}. The latest released Planck cleaned-CMB maps have been produced by the SEVEM template fitting technique, the NILC and SMICA non-parametric procedures and the COMMANDER parametric method. Despite the difference in their approaches, these methods yield largely compatible output maps.

In this work we investigate the application of a new general method for the localization of peaks on the sphere, under isotropic Gaussian noise, to detect point sources out of CMB maps. On one hand, this identification may allow a better cleaning of the CMB maps from the source contaminants; on the other hand, these sources are of astrophysical interest themselves. Here, we focus on intensity CMB maps and leave the extension to polarization to a forthcoming paper. Our method is based on the so-called STEM procedure \citep[see][]{cheng2016}, which consists of the following four steps: i) filtering the CMB maps with (Mexican) needlets, in order to increase the signal-to-noise ratio \citep[see][]{marinucci2008}; ii) detection of local maxima in the filtered maps; iii) computation of $p$-values for each of the local maxima; iv) implementation of the multiple testing scheme by applying the so-called Benjamini-Hochberg (BH) procedure. The new algorithm allows to effectively control the False Discovery Rate, i.e. the expected number of false discoveries among the critical points identified as point sources. We validate our procedure on realistic simulations of CMB maps and, then, apply the algorithm to the latest release of Planck cleaned-CMB maps; in particular, we compare the performance of the different component separation methods in the removal of the extragalactic sources. We find some candidate point sources in most of the maps; however, we conclude that the inpainting procedure adopted for the 2018 release seems to have produced maps which are much closer to being purely Gaussian, with a number of local maxima largely consistent with theoretical predictions.

This paper is organized as follows: in Section 2 we provide a formal description of STEM procedure; in Section 3 we present the numerical implementation of the algorithm and discuss about its performance on simulations; in Section 4 we show our results from Planck cleaned-CMB maps; finally, in Section 5 we draw our conclusions and outline some future perspectives for the application of the algorithm.


\section{Methodology: The STEM Procedure and FDR Control}

The procedure that we shall exploit in this paper can be viewed as an extension to the sphere of the STEM algorithm, which was introduced in a 1-dimensional Euclidean framework by \cite{schwartzman2011}; for the sphere, the mathematical background for our proposal and some theoretical results which we shall present below have been discussed in \cite{cheng2016}.

In short, the algorithm can be summarized in the following four steps (STEM stands for Smoothing and TEsting Multiple hypothesis):

\begin{itemize}
	
	\item In the first step, the map is smoothed to enhance the signal-to-noise ratio of possible sources, and (equivalently) to get rid of as much Cosmic Variance as possible. The proper implementation of this smoothing step is one of the most delicate parts of our algorithm, and is achieved by means of the (Mexican) needlet transform, which we shall describe extensively in Subsection \ref{Mexican}
	
	\item  In the second step, candidate point sources are selected by numerically computing local maxima of the filtered maps. The algorithm to detect the maxima is described in Subsection \ref{Peaks} and has been extensively validated.
	
	\item The third step requires the computation of \emph{p-}values for each of the computed local maxima. Needless to say, even for a purely Gaussian map the distribution of local maxima is not Gaussian, and can be derived analytically along the lines of \cite{bardeen1985,cheng2015} and \cite{cheng2016} (see also \cite{cammarota2016} for related results in the case of single multipole fields/spherical harmonic components). More remarkably, it can be shown that in the high-frequency limit the sample distribution on filtered maps converges to the theoretical expectation, thus making a principled statistical analysis doable. These results are discussed below in Subsection \ref{P-values}.

	\item In the fourth and final step, the multiple testing procedure is implemented. Here we are resorting to the control of the False Discovery Rate, an approach which has become classical in the statistical community over the last decade or so \cite[see][for the pioneering contribution]{benjamini1995}. Heuristically, the idea is to control the proportion of detected point sources that can turn out to be false, as opposed to the control of each of them individually. The advantage of this joint approach to testing have been now very widely recognized in the statistical/mathematical literature, but their impact in Cosmology and Astrophysics has so far been rather limited. The details of our methods, in particular the so-called Benjamini-Hochberg procedure, are discussed below in Subsection \ref{Multiple}.

\end{itemize}

In the subsections to follow, we describe in greater details each of the four steps in our algorithm.

\subsection{(Mexican) Needlets Filtering} \label{Mexican}

The first step in our procedure is the proper filtering of CMB maps in order to enhance the signal-to-noise ratio. Heuristically, point sources are clearly confined to small scales / high frequencies, hence all the features related to the smallest values of the multipoles $\ell$ should be considered as ``noise'' and hence discarded. We are hence looking for a high-pass filter with optimal characteristics.

Needlets are a form of spherical wavelets which were introduced in Cosmology roughly one decade ago and have hence been shown to enjoy a number of very important features. Let us denote by $j, j=1,2,...$ a set of integer-valued frequencies, and by $P_{\ell}(\langle x,y \rangle)$ the family of Legendre polynomials, which for $x,y \in S^2$ satisfies the identity $\frac{2 \ell+1}{4 \pi}P_{\ell}(\langle x,y \rangle) = \sum_{m=-\ell}^{\ell} Y_{\ell m}(x)\overline{Y}_{\ell m}(y)$, where the bar denotes complex conjugation and $Y_{\ell m}$, as usual, the standard basis of spherical harmonics. The needlet filter is then defined to be

\begin{equation}
\psi_j(x,\xi):= \sum_{m=-\ell}^{\ell}b(\frac{\ell}{B^j})\frac{2 \ell +1}{4 \pi}P_{\ell}\langle x,\xi \rangle .
\end{equation}

\begin{figure}
	\centering
	\includegraphics[width=\linewidth]{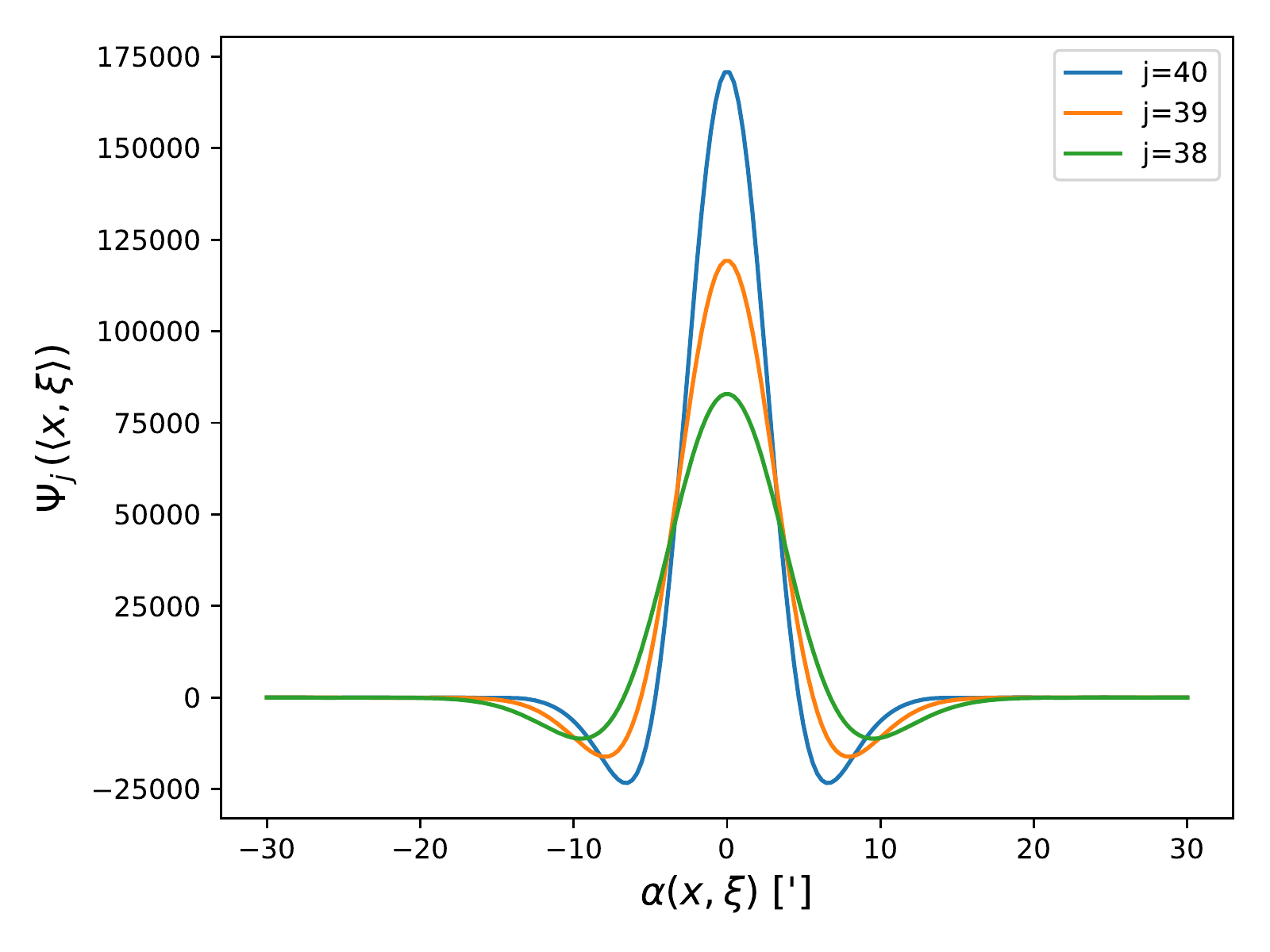}
	\caption{Transversal cut of a needlet function. In the x-axis, it is represented the angular distance between the center of the given needlet ($\xi$) and an arbitrary point $x$, in minutes. The amplitude of the needlet is found on the y-axis. The width of the desired needlet can be selected through the parameters $j$ and $B$. In this work we will fix $B=1.2$ and use the values of $j$ found in the image: $38$, $39$, and $40$. \label{f:pixneedlet}}
\end{figure}

An example of this needlet filter can be seen, represented in pixel space, in Fig.\ref{f:pixneedlet}, with a fixed value of $B=1.2$ and $j$ taking the values $38,39,40$, as we will do throughout this work.

Loosely speaking, the needlet filter is hence nothing more than a weighted average of the usual Legendre polynomial, the latter projecting a spherical map into its distinct multipole components $\ell$; the standard needlet construction was introduced in mathematics by \cite{narcowich2006}, and then in statistics/cosmology by \cite{baldi2009},\cite{marinucci2008}, \cite{pietrobon2006}. The key ingredient in the construction is then the choice of the weight function $b(.)$, which allows the optimal tradeoff between localization properties in the real domain and those in multipole space/frequency domain. In particular, in the standard needlet construction the function $b(.)$ is supposed to be infinitely differentiable, compactly supported, and satisfying the partition of unity property, which entails $\sum_{j}b^2(\frac{\ell}{B^j})=1$; here $B$ is a user-chosen parameter, whose value will be discussed later \cite[see also][]{marinucci2011}. These properties ensure, in particular, that each needlet component is finitely supported in multipole space, and hence very-well localized in frequency space. More than that, it has been possible to show \cite[see][]{narcowich2006} that the needlet filter enjoys very good localization properties in real space, the tail of the filter decaying ``nearly-exponentially", i.e., faster than any polynomial, as the frequency $j$ increases; more precisely, one has that, for all integers $M$ there exists a constant $C_M$ such that

\begin{equation}
\psi(x,\xi) \leq \frac{C_M \times B^{2j}}{(1+B^j \times d_S^2(x,\xi))^M},
\end{equation}
where $d_S^2(x,\xi)$ is the usual geodesic distance on the sphere. The needlet components are then given by
\begin{equation}\label{eq:beta}
\beta_j(\xi):= \int_S^2 T(x) \psi_j(x,\xi)dx = \sum_{\ell}b(\frac{\ell}{B^j})a_{\ell m} Y_{\ell m}(\xi),
\end{equation}
so that on one hand the filtered fields is only supported on the multipoles where the function $b(.)$ takes a non-zero value, while on the other hand the value of the filtered field at any given location $\xi$ is only influenced by the points nearby in the original map $T(x)$. Filtered maps enjoy further very useful (and rather remarkable) properties from the statistical point of view; indeed, it can be shown that for any two different locations in the sky $x$ and $y$, the values of $\beta_j(x)$ and $\beta_j(y)$ are asymptotically uncorrelated \cite[and hence independent, under Gaussianity, see][]{baldi2009,marinucci2008} as the frequency $j$ diverges to infinity. This property will play a crucial role in the determination of the statistical properties of the STEM algorithm.

\begin{figure}
	\centering
	\includegraphics[width=\linewidth]{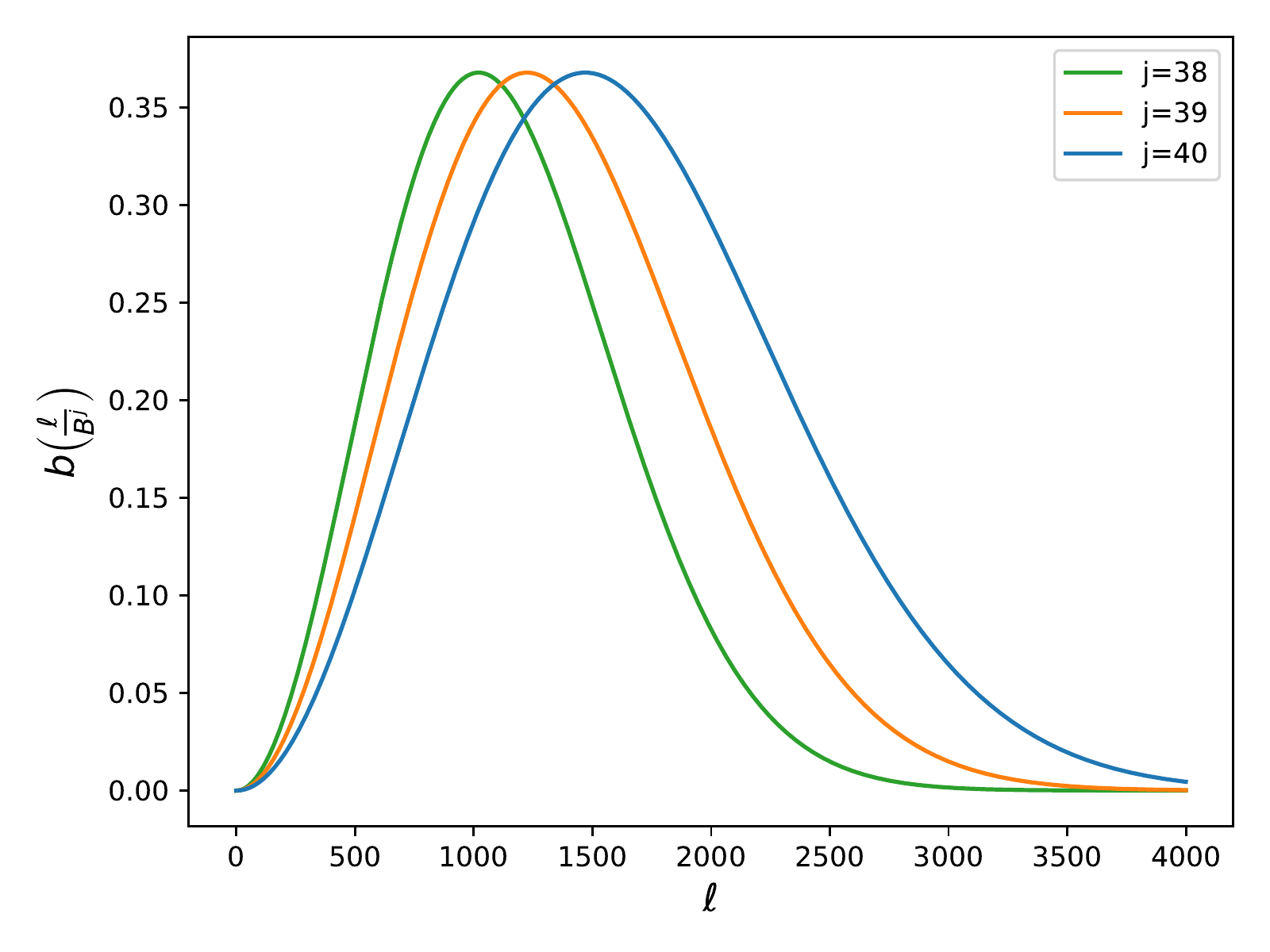}
	\caption{Representation of the filter function $b$ for a Mexican needlet with $B=1.2$ and $j=38,39,40$. The value of $j$ determines the multipole region that the filtering will extract. \label{f:mxneed}}
	\end{figure}

The filter that is actually going to be implemented is indeed a modification of the original needlet idea, which was introduced soon after by \cite{geller2009a,geller2009} and then in the statistical/cosmological literature by \cite{lan2008} , \cite{mayeli2010} and \cite{scodeller2011}. The idea is simply to replace the compactly supported function $b(.)$ by means of the Gaussian-related weight

\begin{equation}\label{eq:mexb}
b_p(\frac{\ell}{B^j}) = \frac{1}{\sqrt{2 \pi}}(\frac{\ell}{B^j})^{2p}exp(-\frac{1}{2}(\frac{\ell}{B^j})^2),
\end{equation}

where $p$ is a parameter, that we have fixed at $p=1$ in what follows, see Fig. \ref{f:mxneed}.

The properties are very much the same as for standard needlets
, however, in some sense Mexican needlets achieve the optimal tradeoff between localization in real and harmonic space. Indeed, their localization properties in the real domain are even better than for standard needlets, as their tails are Gaussian; in harmonic space they are no longer compactly supported on a finite number of multipoles, but for all practical purposes their localization is extremely good, as even in these domain the tails have Gaussian decay.

\subsection{Selection of Candidate Point Sources} \label{Peaks}

Candidate point sources are selected by simply collecting the local maxima in the filtered maps, i.e., the points where the gradient is zero and the (covariant) Hessian matrix is negative definite.
We write $G_(\beta_{j})$ for the set of detected peaks and $M(\beta_{j})$  for their total number, that is to say that a point $x \in S^2$ belongs to $\tilde{G}_{j}$ if and only if
\begin{equation}
\{x \in G(\beta_{j}) \}  \Leftrightarrow \{x: \nabla \beta_j (x)=0 \text{ and } \nabla^2 \beta_j(x) \prec 0\},
\end{equation}
$A \prec 0$ denoting a negative definite matrix.

In practice, candidate peaks are simply identified by the routine \emph{Hotspot} in HEALPix, see \href{https://healpix.jpl.nasa.gov/html/facilitiesnode8.htm}{https://healpix.jpl. nasa.gov/html/facilitiesnode8.htm}.

\subsection{Distribution of Local Maxima and \emph{p-}values} \label{P-values}

The key technical step in our algorithm is based on the possibility to evaluate the exact distribution of local maxima for the CMB maps, under the (null) assumption that no point source is present. Computing the density of local maxima, or equivalently the expected number of maxima for a given Gaussian field, is a topic which has drawn a lot of work in Cosmology, starting from the seminal paper by Bardeen et al. \cite{bardeen1985} in the eighties. Under the circumstances of the present paper, the density of these maxima can be shown to be given by \cite[see][and the references therein]{cheng2016}

\begin{equation}
\label{eq:f}
f_j(x) =\frac{2\sqrt{3+\eta_j^2}}{2+\eta_j^2\sqrt{3+\eta_j^2}}\left[A+B+C \right]
\end{equation}
where we defined
\begin{equation*}
A=\left[\eta_j^2+\kappa_j^2(x^{2}-1)\right] \phi (x)\Phi \left( \frac{\kappa_j x}{%
	\sqrt{2+\eta_j^2-\kappa_j^2}}\right)
\end{equation*}
\begin{equation*}
B=\frac{\kappa_j\sqrt{(2+\eta_j^2-\kappa_j^2)}}{2\pi }xe^{-\frac{%
		(2+\eta_j^2)x^{2}}{2(2+\eta_j^2-\kappa_j^2)}}
\end{equation*}
\begin{equation*}
C=\frac{\sqrt{2}}{\sqrt{\pi (3+\eta_j^2-\kappa_j^2)}}e^{-\frac{%
		(3+\eta_j^2)x^{2}}{2(3+\eta_j^2-\kappa_j^2)}}
\end{equation*}
\begin{equation*}
\times \Phi \left( \frac{\kappa_j x}{%
	\sqrt{(2+\eta_j^2-\kappa_j^2)(3+\eta_j^2-\kappa_j^2)}}\right) \Bigg\}.
\end{equation*}

The notation here requires some clarification. Indeed, $\phi, \Phi$ denote, respectively, the standard density and cumulative distribution function for a Gaussian random variable; on the other hand, $\kappa_j$ and $\eta_j$ are constants which can be explicitly computed from the angular power spectrum of the original (unfiltered) CMB map; more precisely, they are given by
\begin{equation}
\eta_j=\frac{\sqrt{\Gamma_j^{\prime}({C_{\ell }})}}{\sqrt{\Gamma_j^{\prime \prime }({C_{\ell }})}},\qquad \kappa_j=\frac{\Gamma_j^{\prime}({C_{\ell }})}{
	\sqrt{\Gamma_j^{\prime \prime }({C_{\ell }})}},  \label{eq:kappa}
\end{equation}
where
\begin{equation}
\Gamma_j^{\prime}({C_{\ell }}):=\sum_{\ell}\frac{2\ell+1}{4 \pi}b_p(\frac{\ell}{B^j}) P^{\prime}_{\ell}(1),\\
\end{equation}
\begin{equation}
\Gamma_j^{\prime \prime}({C_{\ell }}):=\sum_{\ell}\frac{2\ell+1}{4 \pi}b_p(\frac{\ell}{B^j}) P^{\prime \prime}_{\ell}(1),
\end{equation}
and the derivatives of Legendre polynomials evaluated at $1$ are given by
\begin{equation}
P_{\ell }^{\prime }(1)=\frac{\ell (\ell +1)}{2} \quad\text{ and }\quad
P_{\ell }^{\prime \prime }(1)=\frac{\ell (\ell -1)(\ell +1)(\ell +2)}{8}.
\end{equation}

In the sequel, it should be kept in mind that all our computations will be carried over components which have been normalized to have unit variance.

Of course, once the density of local maxima is known, it is immediate to compute the \emph{p-}value of any one of them taking value $u$ (say), which is indeed given by

\begin{equation}\label{eq:pval}
\pi_j(u):=\int_{u}^{\infty}f_j(t)dt
\end{equation}

A further, more remarkable result was also established in \cite{cheng2016}; it was indeed shown that at high frequencies, the realized (observed) distributions of critical points on CMB maps converges to the theoretical density $f_j$ given above. In other words, even on a single realization of CMB maps the observed distribution of local maxima for large values of $j$ is going to track closely the theoretical prediction, under the assumptions of Gaussianity and isotropy. This result is grounded on the capacity of needlet filtered maps to control Cosmic Variance, and it is clearly the foundation for the statistical multiple testing procedure which we shall describe in the next subsection; see Fig. \ref{f:maxima} later for numerical evidence supporting these claims.

\subsection{The Multiple Testing Procedure} \label{Multiple}

As a multiple testing procedure in step (4) we apply the Benjamini-Hochberg (BH) procedure \cite{benjamini1995}. This procedure is now very popular among statisticians; in this paper
it is implemented directly as follows. Recall that we write $G_(\beta_{j})$ for the set of detected peaks and $M(\beta_{j})$  for their total number; let us arrange their \emph{p}-values in increasing order and we fix a significance level $\alpha \in (0,1)$ whose role will be made clearer below.

\begin{figure}
	\centering
	\includegraphics[width=\linewidth]{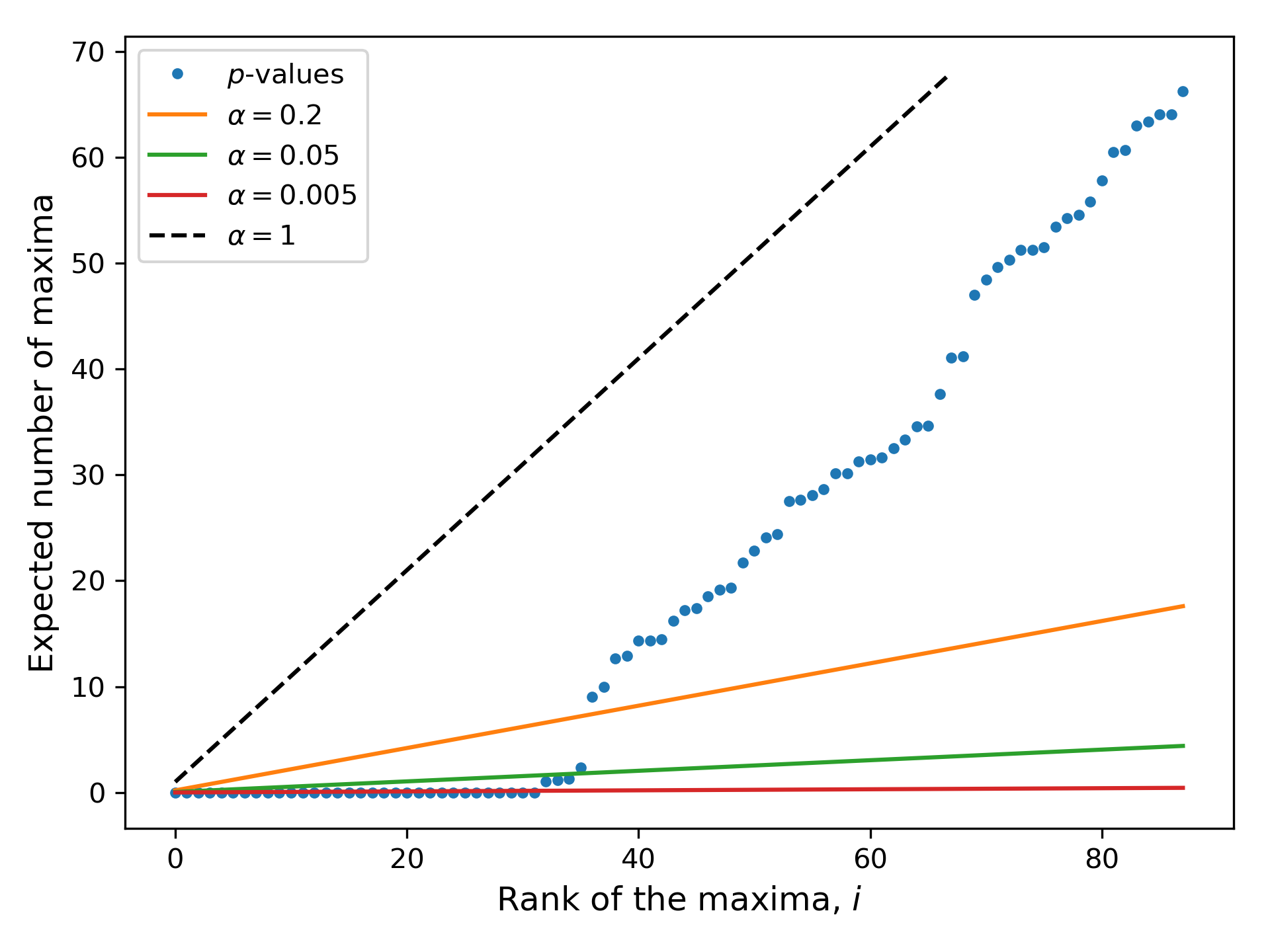}
	\caption{Example of the Benjamini-Hochberg procedure for multiple testing. In the x-axis, we have the rank $i$ of the \emph{p-}values, sorted from lower (less likely) to higher (more likely). The value of the \emph{p-}values times the total number of maxima is plotted for each maximum as blue points. This is an estimate of the number of maxima expected to be at least as intense as a given point in a purely Gaussian CMB map. The three colored lines represent the threshold for 3 different values of $\alpha$; the highest value still below the line determines the number of points to be classified as candidate point sources. In dashed black line, corresponding to $\alpha=1$, we have the expected behaviour for a purely Gaussian map.\label{f:BHpro}
			}
\end{figure}

Now let $k$ be the largest index for which the $i$th smallest \emph{p}-value is less than $i\alpha /M(\beta_{j})$; then the null hypothesis that there is no point source at a given local maximum location $x\in G_(\beta_{j})$ is rejected if
\begin{equation}
p_{j}(x)<\frac{k\alpha }{M(\beta_{j})} ,
\label{eq:thresh-BH-random}
\end{equation}%
where we assume without loss of generality that at least a peak is detected in the map (otherwise the test is clearly unnecessary). In words, the procedure can be explained as follows: we draw a line starting from the origin with slope $\alpha$, and on the same plot we represent in ascending order the \emph{p-}values corresponding to the detected peaks: local maxima are considered significant (and hence identified with point sources) if and only they are small enough to fall below the line. The procedure is illustrated in Fig. \ref{f:BHpro}.

This procedure has a multiple flavour in an obvious sense: for any local maxima, not only its value is considered, but also its ranking in the full map (a single source at $3\sigma$, say, may not be significant in a map with million pixels, but one thousand such maxima have definitely another meaning). More rigorously, this procedure guarantees control of the so called \emph{False Discovery Rate}, which is defined as the expected number of false discoveries out of the critical points which are identified as point sources (see \cite{cheng2016} for more discussion and details):

\begin{equation}
FDR_j=\langle \frac{W_j}{S_j} \rangle ,
\end{equation}
where $S_J,W_j$ denote, respectively, the number of local maxima which are identified as sources and those for which this identification is actually wrong, and $\langle.\rangle$ is as usual the ensemble expected value. Under some standard assumptions, it is indeed possible to show that the STEM algorithm that we introduced on one hand allows to control the False Discovery Rate by the user chosen parameter $\alpha$, in the high-frequency limit; i.e.,
\begin{equation}
lim_{j \rightarrow \infty} FDR_j \leq \alpha ;
\end{equation}
moreover, in a suitable sense the procedure has statistical power growing to unity at the largest scales, i.e., it is able to recover a proportion growing to $100\%$ of the existing sources. These results are clearly of a theoretical nature, but as we shall show in the Sections to follow they do provide a very good guidance on the actual performance of the STEM procedure under realistic experimental conditions and on Planck 2018 data.

\section{Numerical Implementation and Simulations}
The algorithm described in the previous section has been implemented on Planck-like simulations including point sources, as described below.

\subsection{Implementation}

We implement the algorithm by creating a software on Python~$3.5$, exploiting in particular \texttt{numpy} \citep{oliphant2006} and \texttt{scipy} \citep{jones2001}. We use \texttt{astropy} \citep{theastropycollaboration2018} to import \texttt{FITS} images as the ones provided by the Planck Collaboration, and \texttt{HEALPix} \citep{gorski2005} to deal with full-sky spherical images, as this is the standard adopted by the Planck Collaboration. The needlet treatment and multiple testing algorithm explained in the previous section has been programmed entirely by us. Finally, we use \texttt{pandas} \citep{mckinney2010} and \texttt{Matplotlib} \citep{hunter2007} to analyze the results.

The software is structured into three parts: i) The main algorithm, which takes a CMB map and extract a list of points believed to be point sources, according to the STEM Procedure explained in Section 2; ii) The simulations, which create a number of possible CMB realizations and introduces artificial point sources according to specifications, in order to test the algorithm; and iii) The Planck maps analysis, which helps to perform the study of the maps provided by the Planck Collaboration.

Here we explain the main steps of our implementation for each of these parts.

\subsubsection{Implementation of the Main Algorithm}

We follow the four steps described in Section 2: filtering the map, selecting candidates, computing their $p$-values, and applying the multiple testing procedure.

\begin{figure}
	\centering
	\includegraphics[width=0.49\linewidth]{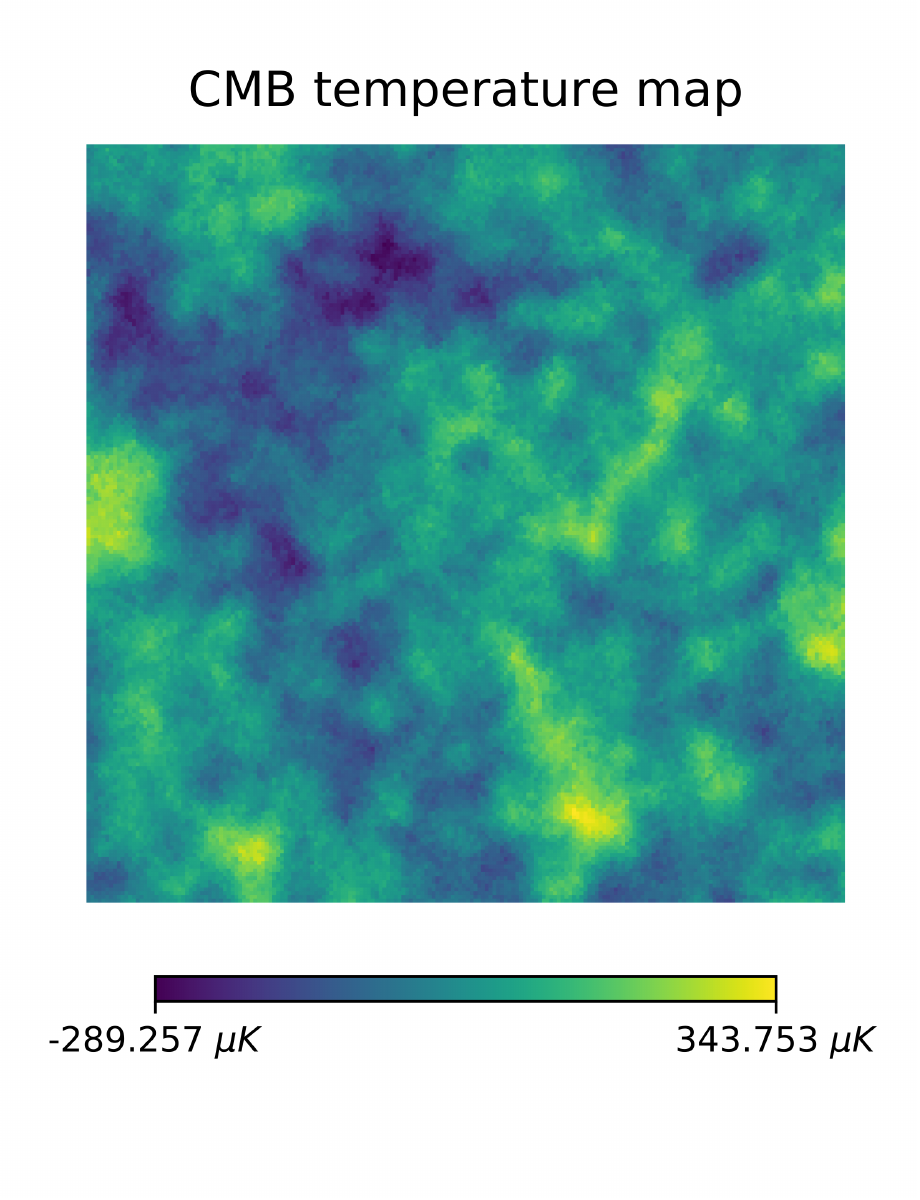}
	\includegraphics[width=0.49\linewidth]{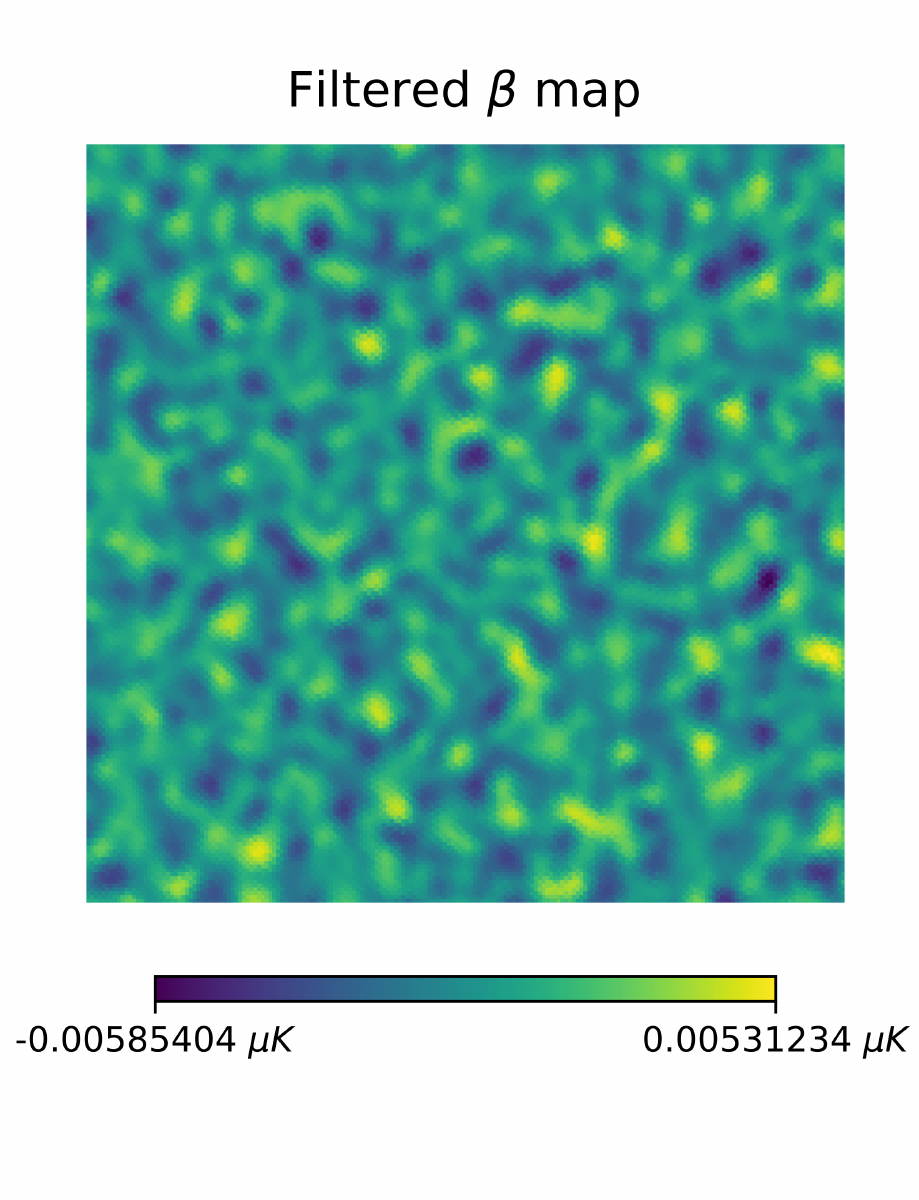}
	\caption{Comparison between a CMB temperature map (on the left) and the $\beta$-map resulting of filtering this map with a needlet $B=1.2$, $j=39$ (on the right). The field of view for both images corresponds to $5\degree\times5\degree$.\label{f:cmb-beta}}
\end{figure}

We start by filtering the input CMB map; this is done in harmonic space, using Eq. \ref{eq:beta}. First, we extract the spherical harmonics coefficients $\alpha_{\ell m}$ using the \texttt{map2alm} routine on \texttt{HEALPix}. Then, these quantities are multiplied by the filter $b \left(\frac{\ell}{B^j}\right)$ of the corresponding Mexican needlet with fixed parameters $B$ and $j$: of course, this filtering factor depends on $\ell$ but not $m$ (needlets are isotropic); this procedure is significantly more efficient than filtering in pixel space. The result is then used to reconstruct the now-filtered map $\beta_{j}(\xi)$ with the \texttt{alm2map} routine on \texttt{HEALPix}.  In Fig. \ref{f:cmb-beta} we can see an example of a CMB map before and after the needlet filtering in a $5'\times5'$ patch of the sky; of course, the small scale clumps in the temperature map are enhanced in the $\beta$-map, especially at the scale corresponding to the selected needlet (around $\sim10''$).

Once the filtered map $\beta_{j}(\xi)$ is computed, we extract its local maxima. This is done with the \texttt{hotspot} routine in \texttt{HEALPix}, which checks the intensity of the neighboring pixels for every pixel in the map (both the location and the intensity of the maxima are delivered). At this point, the maxima distribution can be compared with the theoretical distribution $f_j(x)$.

At this stage, the $p$-value is computed for each maximum, following Eq. \ref{eq:pval}; for each maximum of intensity $x$, we integrate $f_j(x)$ numerically from $x$ to infinity, obtaining the $p$-value $p_j(x)$. However, we are treating tens of thousands of points in each map, and the values of the intensity are similar for a large number of them, so performing this integration for every point is highly inefficient. Instead, we take the highest and lowest values for the intensity and integrate in this interval with a small step of $\Delta x=0.05$; the error in this procedure is lower than the error in the numerical integration, which is $\frac{\Delta p}{p}<10^{-4}$. Additionally, we set the $p$-value to exactly $0$ for extremely high values ($x>37.5\sigma$), in order to avoid computational problems. In the maps we are considering, these intensities correspond to actual $p$-values lower than $10^{-300}$: these values are safely negligible in this context, as they are always reported as point sources.

The last step is to apply the Benjamini-Hochberg procedure to identify the possible point sources among the maxima population. In order to do that, we proceed as described in Section \ref{Multiple}: first we sort the $p$-values from lowest (less likely) to highest (more likely); and then we extract the highest $p$-value that satisfies Eq. \ref{eq:thresh-BH-random}. This routine can be implemented for any value of the confidence parameter $\alpha\in(0,1)$, see Fig. \ref{f:BHpro}.

\subsubsection{Implementation of the simulations}

We are now in the position to generate a number of realizations of the CMB and a set of artificial point sources, according to the input parameters; after applying the STEM algorithm, our routine calculates the true and false detections obtained by the algorithm.

More precisely, we generate an angular power spectrum $C_\ell$ from the Planck cosmological parameters, using \texttt{CAMB} \citep[see][]{lewis2011}. We use this angular power spectrum to generate a series of compatible CMB maps; we choose to use the same resolution of the Planck maps, $\texttt{nside}=2048$. According to the \texttt{HEALPix} standard, the total number of pixels is $12\cdot \texttt{nside}^2$: the maps are then convolved with a Gaussian profile of $fwhm=5'$, similar to the nominal resolution of Planck data.

A population of artificial point sources is then introduced. In order to do that, a blank map of higher resolution is created (\texttt{nside} $=4096$) and a set of pixels are selected at random; the intensities of these points are selected in such a way that this population follows a uniform distribution between the chosen limits. Hence, these artificial point sources are convolved with the same Gaussian profile as used for the map and added to the different CMB realizations.

The main algorithm is then run on each map to obtain the list of points selected as point sources. We will claim that a detection is successful if it is less than $\rho=3$ pixels away from the artificial source that had been injected into the map, as in \cite{cheng2016}. We find that increasing this tolerance to any reasonable extent does not have a significant impact on the results, while of course excluding this tolerance factor decreases the number of detections.

The values chosen for the parameters and the results of these simulations will be explained in Section~\ref{ss:validation}.

\subsubsection{Implementation for the Planck data}

The last part of our numerical implementation concerns a pipeline that loads the target maps (in this case Planck CMB observations), applies the algorithm, and gather the most important results (which we present in the tables below). First, we extract some information concerning each map: number of pixels, map-making algorithm (and its version), and whether it is the inpainted version provided by Planck or not. Then, the user can choose a set of values for the parameters of the main algorithm: in particular, the needlet parameters $B$ and $j$, and the confidence level $\alpha$.

After the main algorithm is implemented on each map, the procedure checks the location of the selected point sources; in order to do that, it uses the two confidence masks provided by the Planck Collaboration. The first one is the common confidence mask, obtained through the combination of the masks for the individual algorithms; it covers around $\sim22\%$ of the sky. The second is the inpainting mask, covering the areas with apparent contamination that are to be inpainted to obtain realistic maps; it covers around $\sim2\%$ of the sky (see Section 4.2 in \cite{planckcollaboration2018a} for more information about the masks). In our implementation, the number of point sources detected inside and outside each mask is reported; the Planck Catalogue of point sources is also explored to count how many of the detections in each region are matched by known sources.

We then produce a \texttt{pandas} table with the results; for every map we can find the information about the parameters of the algorithm and the number of sources reported (total number, number inside each mask and number in the catalogues).

\subsection{Numerical Validation}
\label{ss:validation}

Let us now describe quickly our validating simulations. In the case of Mexican needlets, the only free parameter to be chosen is $B^j$, see Eq. \ref{eq:mexb};  in particular, we select $B=1.2$ at frequencies $j=38, 39, 40$, meaning multipole regions around $B^j \approx 1020, 1225, 1470$. Selecting lower multipoles (larger scales) makes the point source detection less efficient, while selecting higher multipoles (smaller scales) makes the algorithm more sensitive to noise and pixelization effects.

\begin{figure}
	\centering
	\includegraphics[width=\linewidth]{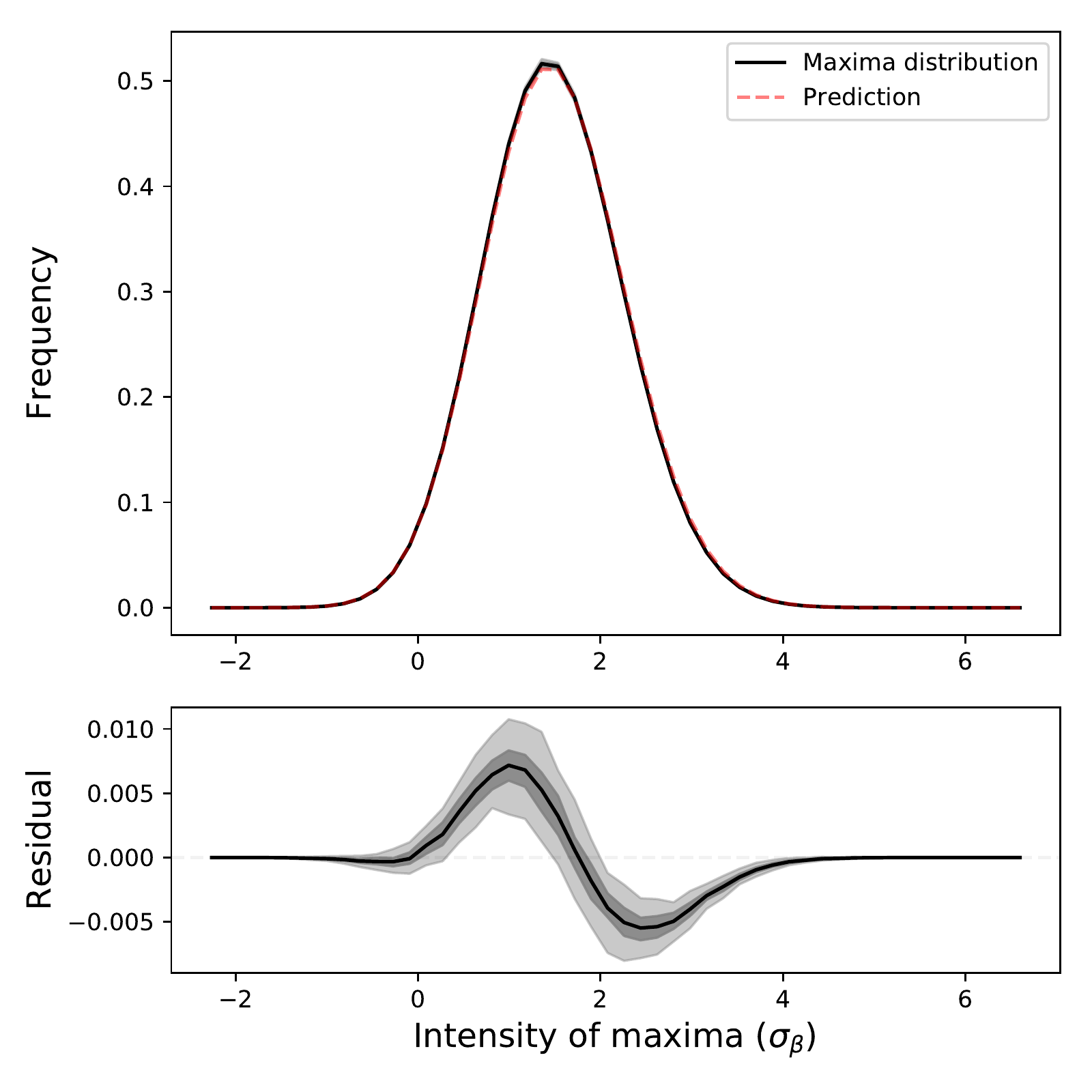}
	\caption{Maxima distribution of filtered CMB maps. On the top, the comparison between the observed distribution (mean of $200$ realizations), in continuous black line; and the theoretical prediction from Eq. \ref{eq:f}, in dashed red line. On the bottom, the residual: $measured-predicted$. Both cases include the regions where the $68\%$ and the $90\%$ of the maps lay.\label{f:maxima}}
\end{figure}

We start by generating $200$ CMB realizations and observing that the maxima population follows the theoretical distribution from Eq. \ref{eq:f}, as it can be seen in Fig. \ref{f:maxima}. We note that the residual is not exactly $0$, but it presents a characteristic pattern, possibly due to pixelization effects. However the case may be, the maximum of the residual is less than $0.01$; this is aligned with the result by \cite{cheng2016}, which showed that at high frequency, the maxima distribution of CMB maps converge to the theoretical prediction.

Using again the sample of $200$ CMB realizations, the algorithm is applied to them with different values of the confidence parameter $\alpha=0.05, 0.01, 0.002$. Since no artificial source is added, we expect a very low amount of reported point sources.

The number of point sources reported can be seen in Table \ref{table:nosources}. In general, the number of detections that are reported will increase as $\alpha$ increases. As expected, this number is $0$ in most of the maps: even for the ``high'' value of $\alpha=0.05$, only $7.5\%$ of the maps report any point source (and only one map reports more than one point).  We select $\alpha=0.01$ as our working standard; however, all the calculations are carried for the three different values since this part of the computation is very efficient.

\begin{table}
	\centering
	\begin{tabular}{ccc}
		\hline
		$\alpha$ & Maps with & Total number\\
		& candidates &  of candidates \\
		\hline  $0.05$ & $15\:/200$ & $17$ \\
		$0.01$ & $7\:/200$ & $7$ \\
		$0.002$ & $1\:/200$ & $1$ \\ \hline
	\end{tabular}\vspace{5pt}
	\caption{Point sources obtained for a total run of $200$ CMB maps without added artificial sources. As expected, very few candidates are reported and the number increases with $\alpha$.\label{table:nosources}}
\end{table}

\subsubsection{Sensitivity of the algorithm}

In order to test the detection power of the algorithm, we introduce a total of $200$ artificial point sources in the simulated maps; they are produced with a peak intensity between $0$ and $7\sigma$, were $\sigma$ is the standard deviation of the convolved maps (calculated from the theoretical angular power spectrum). We are interested in knowing down to what intensity the algorithm is able to recover most point sources: this will be the lower limit on the intensity for which we expect to be complete in the detection. Of course, we are also interested in the number and intensities of false detections.

\begin{figure}
	\centering
	\includegraphics[width=\linewidth]{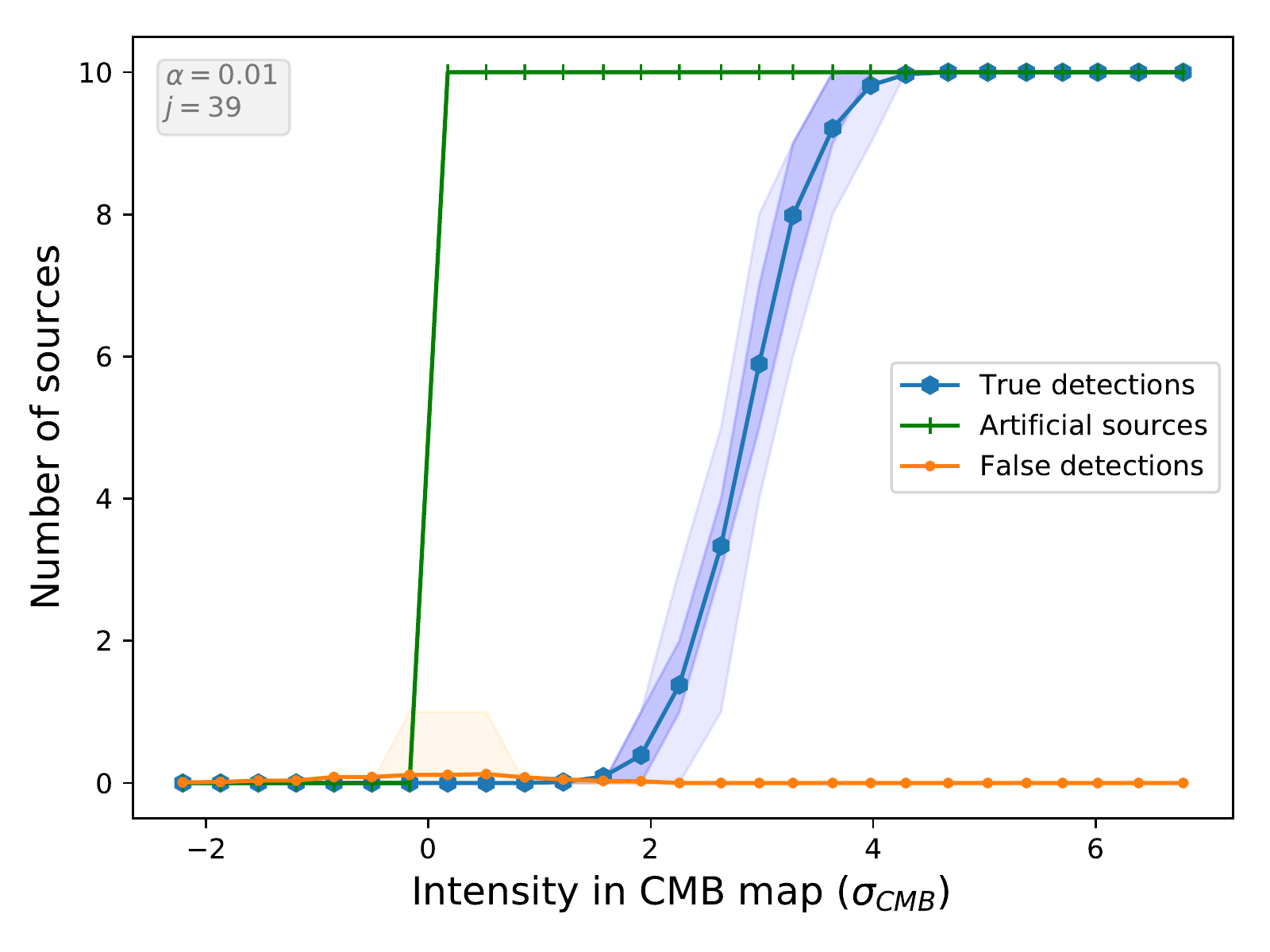}
	\caption{Sensitivity of the algorithm. In the x-axis, we have the intensity of the point sources, referred to the CMB temperature maps before filtering, in units of its standard deviation. In the y-axis, we have the number of detection of the maxima in each bin. In green, we have the artificial sources; 10 sources have been introduced in each bin, between $0$ and $7\sigma$. In blue, we have the true detections: number of artificial sources correctly detected. In orange, we have the false detections: reported candidates that do not correspond to any artificial source. For both true and false detection, we plot the average and the confidence regions which include $70$ and $90\%$ of the maps. The intensity for the artificial sources is the one with which they were generated, the one for the false detections is measured on the map. The algorithm has been applied $200$ times, with $200$ sources at a time, a needlet filter of $B=1.2$ and ${j}=39$, and $\alpha=0.01$.\label{f:sensitivity}}
\end{figure}

The results of the simulations for $\alpha=0.01$, $B=1.2$, and $j=39$ can be seen in Fig. \ref{f:sensitivity}, which provides the number of sources, the correct and the false detections; the intensity of the maxima is measured with respect to the CMB temperature map before the needlet filtering, in units of its standard deviation $\sigma$. It can be seen that the algorithm detects essentially every artificial source over $4\sigma$ and more than half over $3\sigma$; additionally, the number of false detections is very small and almost negligible over $2\sigma$.

\begin{figure}[p]
	\centering
	\includegraphics[width=\linewidth]{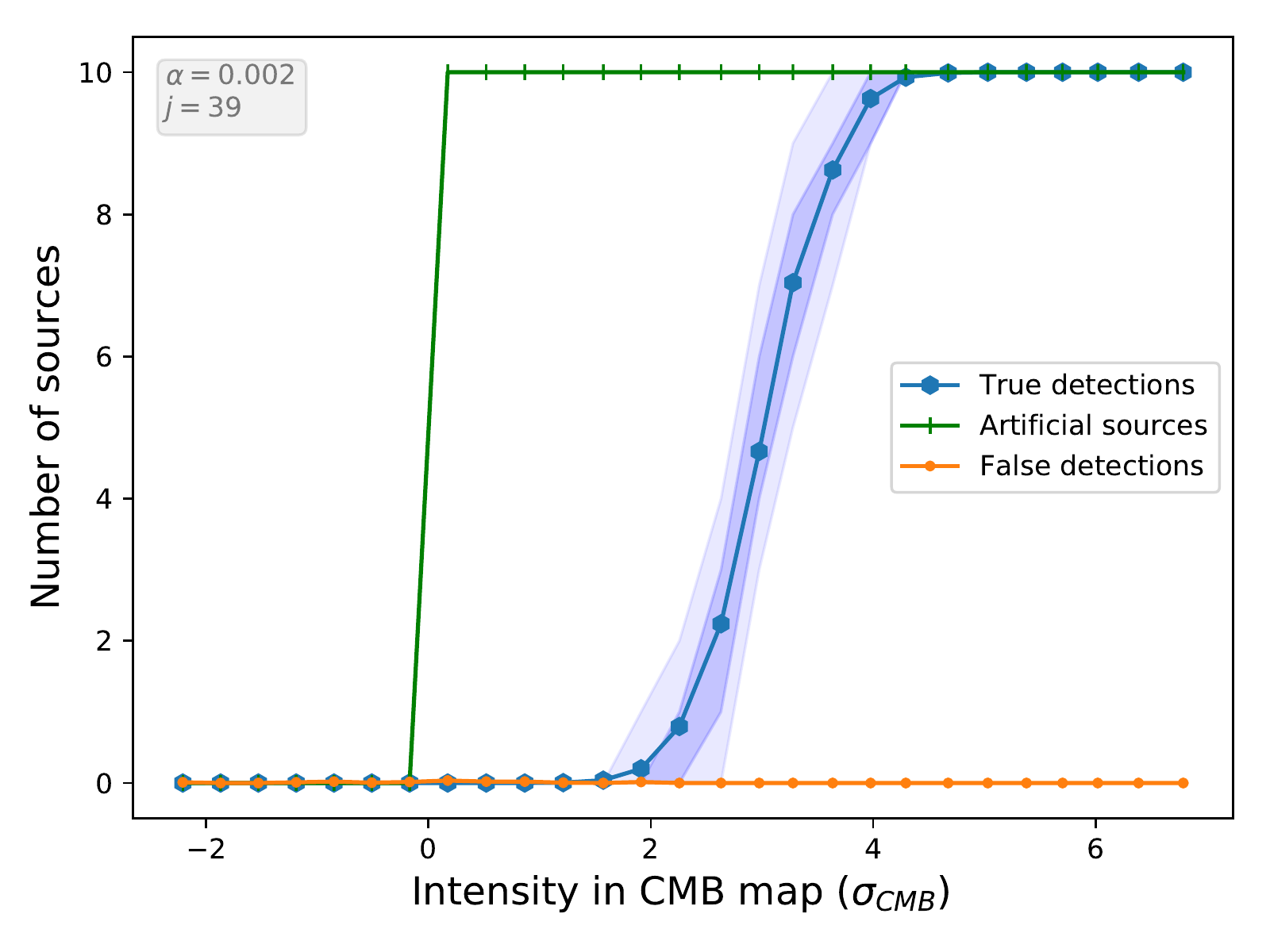}
	\includegraphics[width=\linewidth]{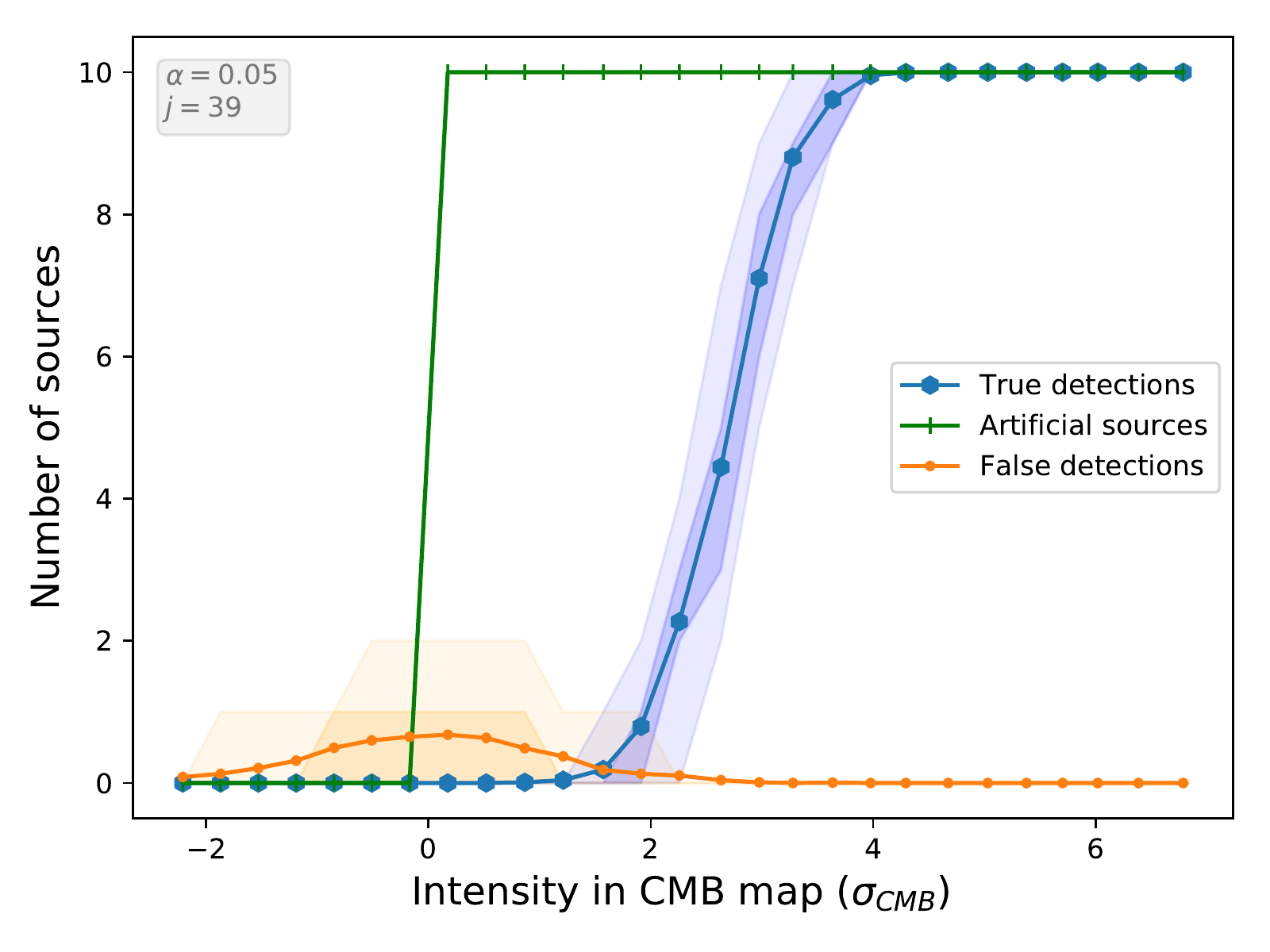}
	\caption{Same as Fig. \ref{f:sensitivity}, with different values for the confidence parameter. On the top, $\alpha=0.002$; on the bottom, $\alpha=0.05$. Higher values include less confident detections, which increases the sensitivity of the algorithm but produces a higher number of false detections.\label{f:sensitivity2}}
\end{figure}

Of course, a lower value of $\alpha=0.002$ implies a higher value of intensity for which almost all signals are detected (around $4.5\sigma$); on the other hand, the number of false detections is reduced to negligible values. Likewise, a higher $\alpha=0.05$ entails that almost all sources above $3\sigma$ are correctly detected, but there is a significant population of false detections. This effect can be seen in Fig. \ref{f:sensitivity2}.

\begin{figure}
	\centering
	\includegraphics[width=\linewidth]{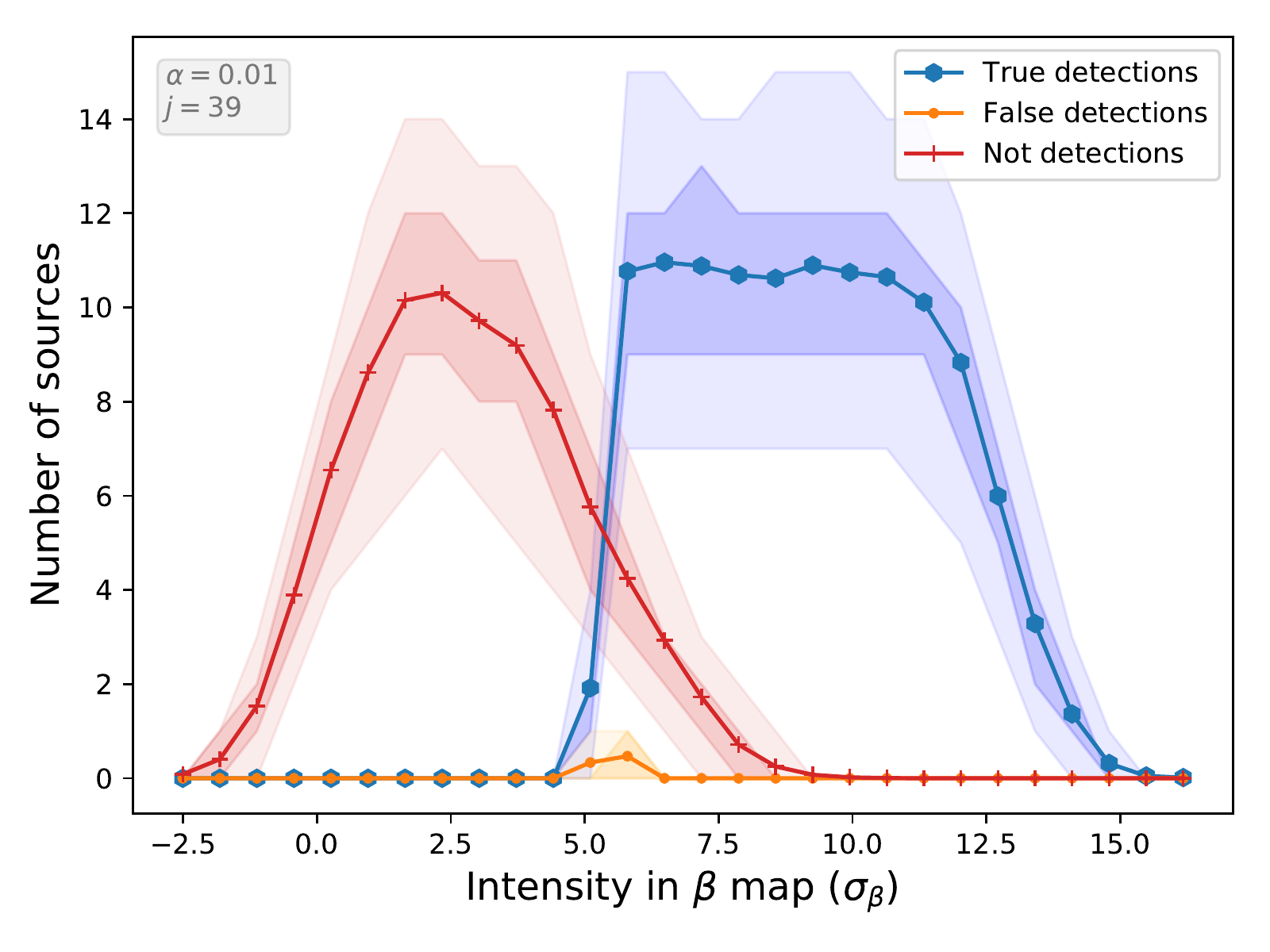}
	\caption{Sensitivity of the algorithm on the filtered CMB maps ($\beta$-maps). In the x-axis, we have the intensity of the point sources, measured directly on the filtered CMB maps. In the y-axis, we have the number of point sources. Detections that correspond to artificial point sources are plotted in blue, while detections that do not, are represented in orange. Artificial point sources that are not detected are represented in red. The average and the confidence regions are plotted, where $70$ and $90\%$ of the maps lay. As before, the algorithm has been applied $200$ times, with $200$ sources at a time, a needlet filter of $B=1.2$ and ${j}=39$, and $\alpha=0.01$.\label{f:sensitivitybeta}}
\end{figure}

The previous results are reported with the intensity referring to the CMB temperature map, before the filtering is applied. Since the detection occurs on the filtered maps $\beta$, it is interesting to see the intensity of the point sources after filtering; this is given in Fig. \ref{f:sensitivitybeta}. The first difference is a clear enhancement of the signal: while point sources are generated with intensities between $0$ and $7\sigma_{CMB}$, the intensity of these points in the $\beta$-maps is up to $15\sigma_\beta$.

Another difference is the noisy profile of the detections; this is because the intensity here is measured on the filtered map after adding CMB, instead of the intensity of the artificial point source by itself.

This plot is obtained for $\alpha=0.05$; lower values imply that a small part of the lower end of the graph is dismissed (the lower cut for the intensity increases): at $\alpha=0.002$, the lowest intensity is around $6\sigma_\beta$. As we saw before, this will significantly reduce the false detections but will also reduce the faintest end of the true detections.

\begin{figure}[p]
	\centering
	\includegraphics[width=\linewidth]{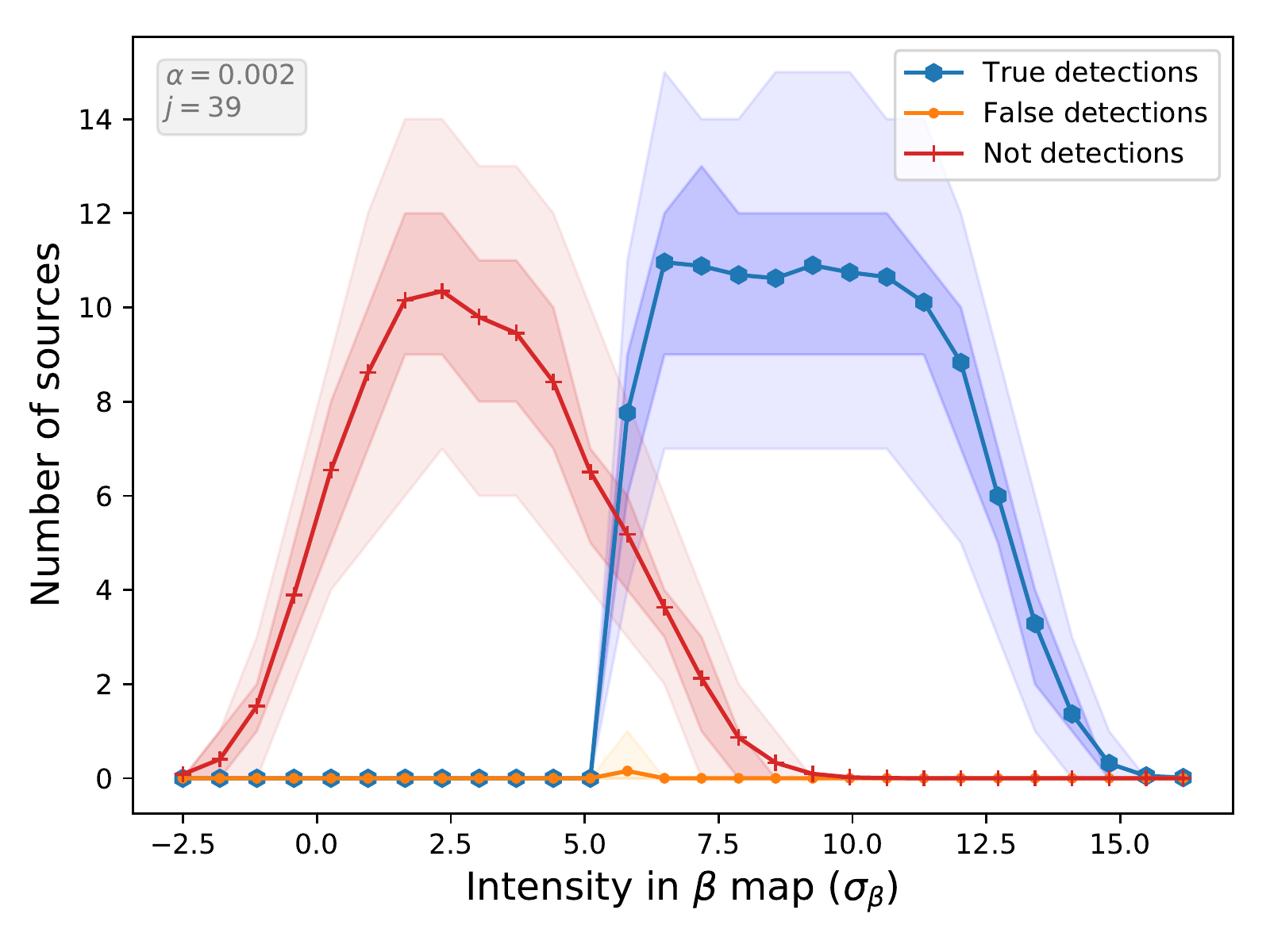}
	\includegraphics[width=\linewidth]{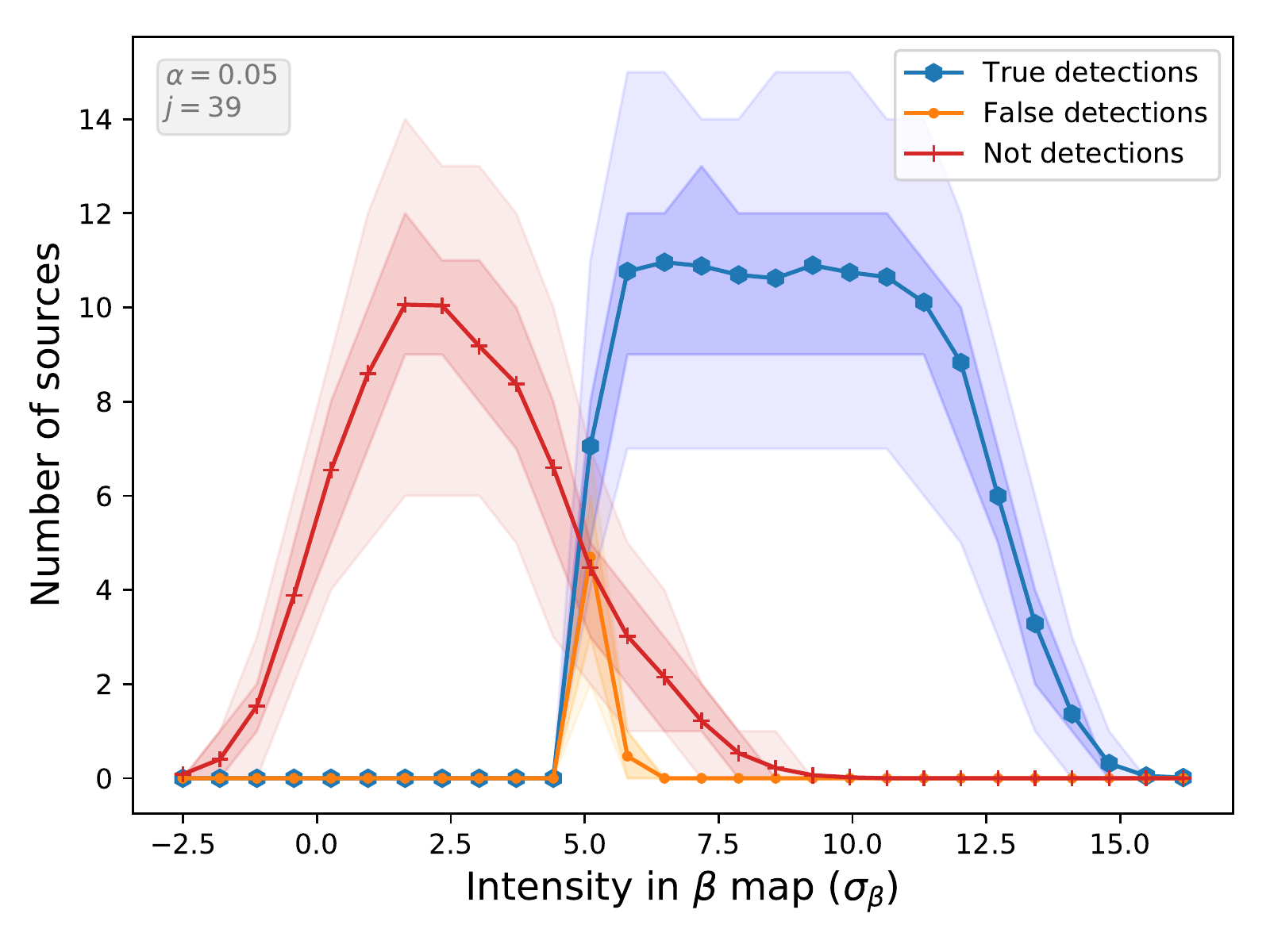}
	\caption{Same as Fig. \ref{f:sensitivitybeta}, with different values for the confident parameter. On the top, $\alpha=0.002$; on the bottom, $\alpha=0.05$. As before, increasing $\alpha$ means that the algorithm successfully detects fainter point sources (blue), but includes more false positives (orange). \label{f:sensitivitybeta2}}
\end{figure}

\subsubsection{Noise and masking}

In the previous sections we have not considered the effects of noise and masking. We are going to explain their effects in this section.

\begin{figure}
	\centering
	\includegraphics[width=\linewidth]{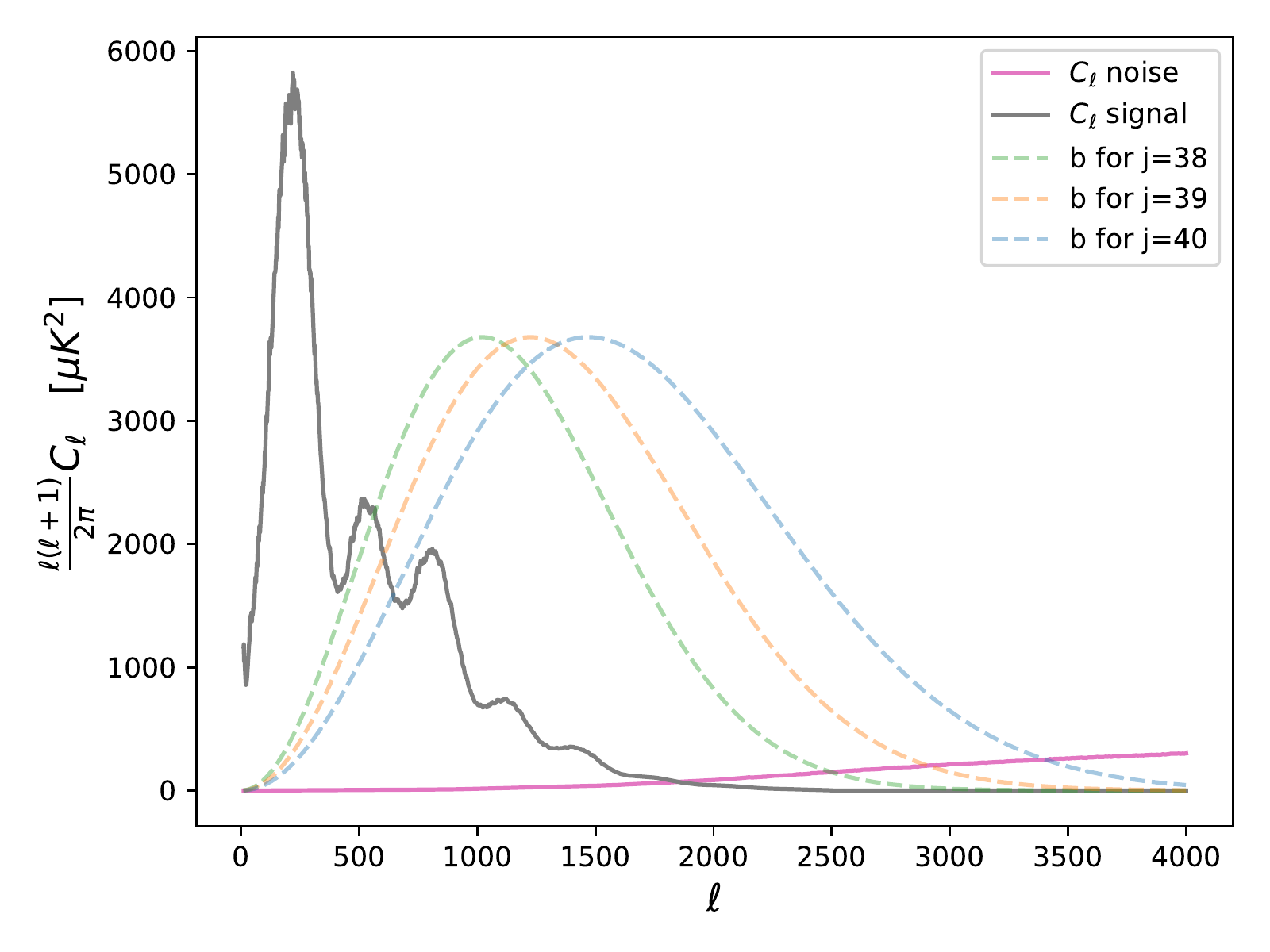}
	\caption{Angular power spectrum $C_\ell$ of the signal and the noise. Overlapped in dashed line, we can see the filter $b\left(\frac{\ell}{B^j}\right)$ for the needlets used in this work: $B=1.2$ and $j=38,39,40$. As expected, low multipoles region is dominated by signal, while high multipoles are dominated by noise. The needlet scales are chosen in order to reflect the highest possible multipoles while staying in the signal-dominated region.\label{f:clnoise}}
\end{figure}

One of the advantages of needlet filtering is that it focuses only on the contribution of a band-limited multipole region; noise contributes more importantly at high $\ell$, so we can minimize its effect by choosing lower values for $\ell$: these scales can be chosen large enough to be still suitable to detect point sources. In Fig. \ref{f:clnoise}, we can see the contribution of the noise and the signal to the angular power spectrum $C_\ell$, which is plotted together with the shape of the needlet filter $b$ for the working values $B=1.2$, $j=38,39,40$. It can be seen that the needlet filtering will reflect only the scales where the noise contribution is less relevant.

We also test this result with simulations, adding a strong noise component to the maps before applying the algorithm. As expected, we do not observe any significant difference in the set of reported point sources.

On the other hand, we have the effects of masking, i.e., setting to $0$ the pixels of areas considered to be contaminated by the Galaxy or known point sources. We recall that one of the main advantages of using needlets is their extremely good localization properties; therefore, we may expect that masking would remove the areas with known contamination, while not affecting the algorithm on the rest on the map. There is, however, some technical difficulties that may arise: since masking a region makes its value equal to $0$, some masked sources in cold (negative) regions can still appear as very intense point sources; additionally, in some cases masking generates artificial detections near the border of the masked area. We avoid these technical problems by using the masks only to check the location of the detections (i.e., inside or outside the confidence masks); we do not mask the map before applying the algorithm.

\section{Results on Planck 2018 Maps}
In this section we apply the algorithm to the temperature maps extracted by the four algorithms of the last Planck data release: COMMANDER, SMICA, SEVEM, and NILC. All the maps are processed at their native resolution of \texttt{nside} $=2048$. We analyze the $2015$ and $2018$ data releases \citep[see][]{planckcollaboration2016e,planckcollaboration2018a}; for the latter, we apply the algorithm both in the inpainted and not inpainted cases. These maps are expected to be very close outside the confidence masks, but the results will be different since the algorithm considers the maxima population as a whole in the entire map. Using inpainting is a way to exclude the bright and known sources from this analysis.

The algorithm is applied with the same parameters as before: $B=1.2$, $j=38,39,40$, and confidence parameter $\alpha=0.05,0.01,0.002$. Several results are stored for each map: first, the total number of detections reported (their location and intensity are also stored but will not be shown here for brevity's sake). We then calculate the amount of the detections outside the inpainting mask ($98\%$ of the sky) and outside the common confidence mask ($78\%$ of the sky). Lastly, we check how many detections match the Planck catalogues of point sources, both in the total map and outside the common confidence mask; these catalogues are reported for each frequency, while here we check all the frequencies at the same time.

The catalogues are checked separately for the $80\%$ confident detection (PCCS catalogues), for where confidence can not be determined (PCCS\textunderscore{}excluded catalogues), and for both together. We note that some sources may be present in both catalogues for different frequencies, especially for the lower frequencies LFI bands, where there is only one catalogue per band.

The file containing the complete table with all the results can be found in \href{http://javiercarron.com/publications}{http://javiercarron.com/publications}. This table consists of $144$ rows and $18$ columns, including the location and intensity of every point reported by the algorithm and, therefore, is too large to be reproduced in this article. In Table \ref{table:res} we can see a part of these results for $j=39$ and $\alpha=0.01$, on which we are going to focus.

\begin{table*}
	\centering
	\centerline{
		\begin{tabular}{lrrrrrrr}
			\hline
        Map & version & inpainted & detections & mask C & mask I & cat & cat C \\ \hline
  commander &    2.01 &     False &         55 &                   0 &                       2 &     53 &           0 \\
  commander &    3.00 &     False &       6335 &                  37 &                     717 &   2927 &          11 \\
  commander &    3.00 &      True &         70 &                   1 &                      12 &      8 &           0 \\
       nilc &    2.01 &     False &       1222 &                   6 &                      10 &    753 &           1 \\
       nilc &    3.00 &     False &       1240 &                   8 &                      15 &    755 &           1 \\
       nilc &    3.00 &      True &          0 &                   0 &                       0 &      0 &           0 \\
      sevem &    2.01 &     False &       7064 &                  39 &                     690 &   4056 &          13 \\
      sevem &    3.00 &     False &       5894 &                  31 &                     261 &   3424 &          10 \\
      sevem &    3.00 &      True &          3 &                   0 &                       1 &      0 &           0 \\
      smica &    2.01 &     False &        558 &                   2 &                      10 &     69 &           0 \\
      smica &    3.00 &     False &        318 &                   2 &                      16 &    197 &           0 \\
      smica &    3.00 &      True &         11 &                   1 &                       3 &      7 &           0 \\ \hline
	\end{tabular}}
	\caption{Part of the results table. Here, the three versions for the four algorithms have been filtered with a needlet $B=1.2$ and $j=39$. Also, $\alpha$ is fixed to $0.01$. The columns are as follows. Map: algorithm used to extract the cleaned temperature map. Version: data release used, $2.01$ corresponding to $2015$ and $3.00$ to $2018$. Detections: number of candidates reported. Mask C: number of candidates \emph{outside} the common confidence mask ($78\%$ of the sky). Mask I: number of candidates \emph{outside} the inpainting mask ($98\%$ of the sky). Cat: number of candidates present in a catalogue. Cat C: number of candidates \emph{outside} the common confidence mask that are present in a catalogue.\label{table:res}}
\end{table*}

We note that the algorithm reports a significant number of detections for most of the maps. Some of them are sources already present in the Planck catalogues that, apparently, could not be completely removed for the final maps. However, a fraction of them are sources that are not intense enough to be present in these catalogues.

Before analyzing the results for the different algorithms, we are going to focus on the similarities and the general behavior of the algorithm. The effect of the frequency filtered ($j$) and the confidence parameter $\alpha$ is similar for all maps. We can see an example of this effect for the not-inpainted 2018 SMICA map in Table \ref{table:jalpha}.

\begin{table*}
	\centering
	\begin{tabular}{lrrrrrrr}
		\hline
    Map & version & inpainted &   j & alpha & detections & mask C & mask I \\ \hline
  smica &    3.00 &     False &  38 &  0.01 &        169 &                   1 &                       6 \\
  smica &    3.00 &     False &  39 &  0.01 &        318 &                   2 &                      16 \\
  smica &    3.00 &     False &  40 &  0.01 &        744 &                   5 &                      79 \\ \hline
  smica &    3.00 &     False &  39 &  0.002 &        268 &                   1 &                       8 \\
  smica &    3.00 &     False &  39 &   0.01 &        318 &                   2 &                      16 \\
  smica &    3.00 &     False &  39 &   0.05 &        403 &                  11 &                      49 \\ \hline
	\end{tabular}
	\caption{Example of the behaviour of the results with $j$ and $\alpha$ for the not-inpainted SMICA maps from the last data release. On the top, varying $j$ with fixed $\alpha$. On the bottom, varying $\alpha$ with fixed $j$.\label{table:jalpha}}
\end{table*}

A lower value of $j=38$ ($\ell\sim1020$) means filtering the map with wider needlets; this means that the signal will be more diluted, making it more difficult to detect sources but more robust against noise. On the other hand, a higher value of $j=40$ ($\ell\sim1470$) means filtering with narrower needlets, which are similar in size to the point sources; this makes this value more sensitive to point sources but also more vulnerable to noise. This explains the observed result: the number of detections grow with $j$.

Similarly, the number of detections also grows with the confidence parameter $\alpha$. As explained before, a higher value of $\alpha$ means that the confidence required to report a detection is lower; therefore, more detections will be reported. With these values for $\alpha$ we are able to detect a significant number of point sources introducing a low count of false detection, as explored with simulations in Section \ref{ss:validation}.

\subsection{Comparison between the different algorithms}

We are now going to analyze the differences between the results for the different algorithms used to extract the CMB temperature map; we are also going to discuss the version (data release of the map and whether it is inpainted or not). We use the results for the parameters $j=39$ and $\alpha = 0.01$ as the standard, but other choices lead to similar trends.

SEVEM maps present the highest amount of point sources for the $2015$ version ($7064$). This number is reduced for the $2018$ version of the map, with $5894$ detections. In these two cases the Galactic component is strong and produces a large number of detections in the Galactic plane; indeed, only a small number of them ($39$ and $31$, respectively) are detected outside the confidence mask. As mentioned before, we also analyze the inpainted map in order to limit the influence of the Galactic contamination; in this case, the overwhelming majority of the detections are removed, only $3$ remaining. None of these detections is outside of the confidence mask: this means that the maxima population of the SEVEM inpainted map is compatible with the purely Gaussian case outside the confidence mask.

NILC maps present a lower amount of detections overall: $1222$ and $1240$ for the $2015$ and $2018$ versions, respectively: again, only a very small part ($6$ and $8$) are outside the confidence mask. More interesting here is the case of the inpainted map: this is the only case where the algorithm does not report any detection, not even inside the masks. Once the strongest point sources are removed through inpainting, the maxima population of the NILC map is perfectly compatible with the Gaussian case.

\begin{figure*}
	\centering
	\includegraphics[width=0.9\linewidth]{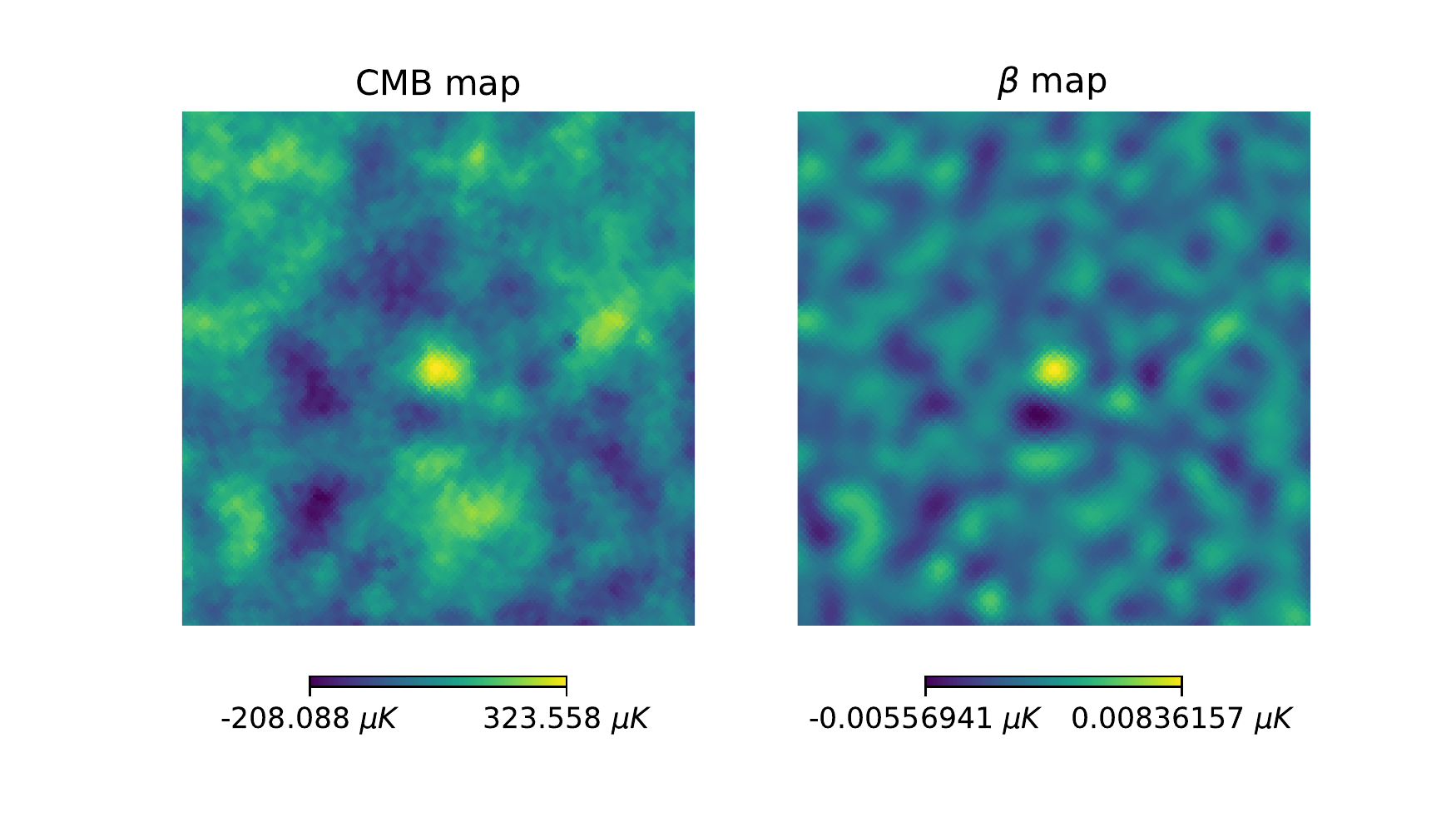}
	\caption{Point source detected by the algorithm in the SMICA map. This point is located at galactic longitude $l=103\degree 18'$ and galactic latitude $b=-27\degree 17'$. The signal to noise ratio at the center is $3.01$ in the CMB map and $5.84$ in the filtered map. It has also been reported in the COMMANDER map, where it has a signal to noise ratio at the center of $2.94$ in the CMB map and $5.69$ in the filtered map. The image corresponds to a $3\degree\times3\degree$ patch of the sky. \label{f:pssmica}}
\end{figure*}

SMICA maps present the lowest number of detections: $558$ and $318$ for the $2015$ and $2018$ versions, respectively. Of these detections, only $2$ are outside the confidence mask in both cases. The inpainted SMICA map present $11$ detections, a higher quantity that the previous algorithms; of these, only one of them is located outside the confidence mask. This source can be seen in Fig. \ref{f:pssmica}.

COMMANDER maps present the highest number of detections in the $2018$ version ($6335$), although it was the lowest for the $2015$ version ($55$). This is probably due to the fact that, in the last version, this is the only algorithm that does not preprocess the frequency maps to remove possible point sources. This produces a more robust procedure in exchange of a higher level of contamination. Indeed, if we look at the inpainted map, where most of these points are removed a posteriori, the number of detections is reduced to $70$, only $1$ of them located outside the confidence mask. This point is the same that was reported in the SMICA map, which can be seen in Fig. \ref{f:pssmica}.


In general, we observe that the inpainting procedures have worked extremely efficiently, and all four algorithms seem basically without or with very few spurious maxima (i.e., undetected point sources) after inpainting has been implemented.

\subsection{Reported point sources}

\begin{figure}
	\centering
	\includegraphics[width=\linewidth]{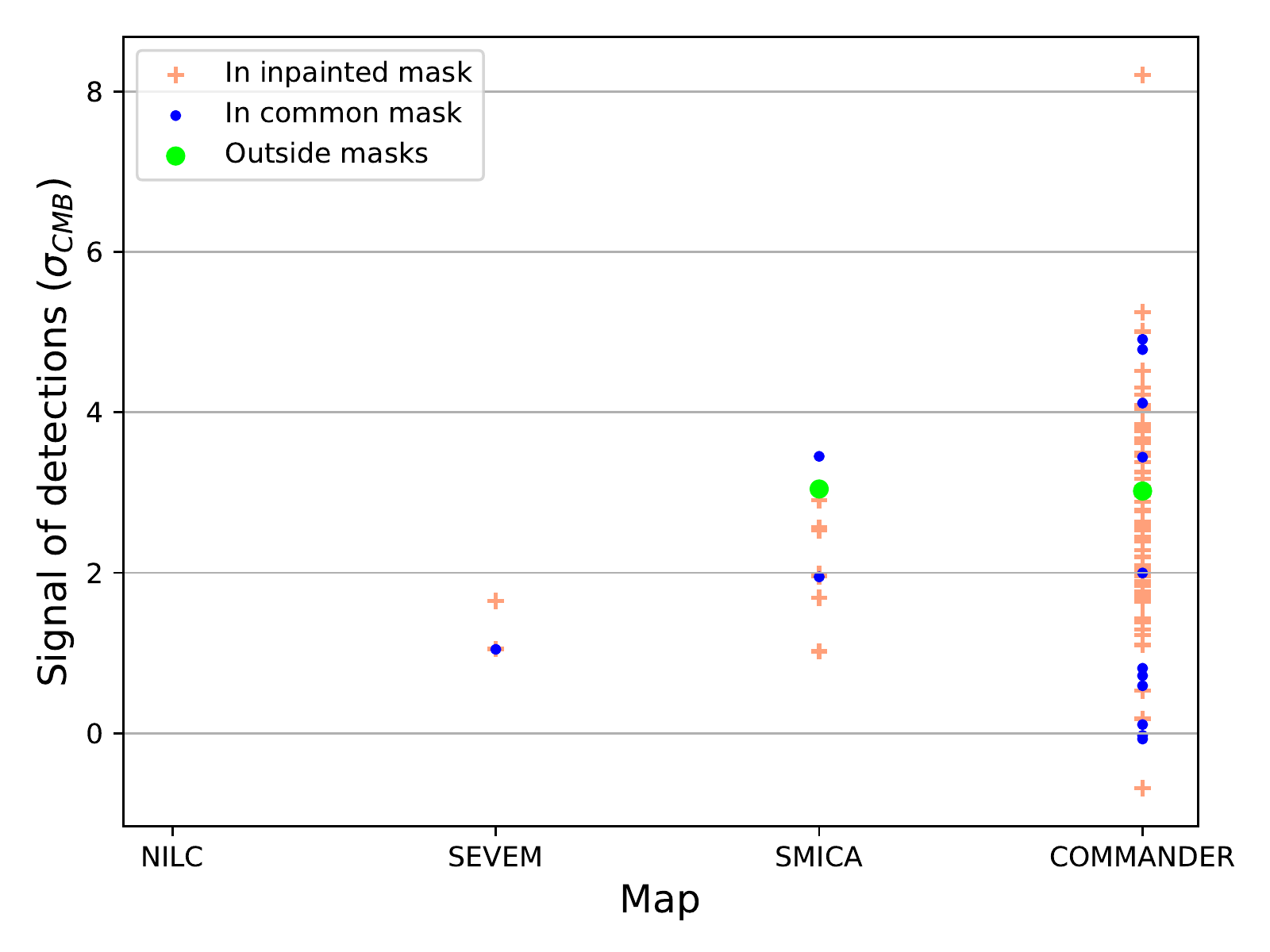}
	\caption{Signal of the detections reported by the algorithm of the impainted maps, measured on the (unfiltered) CMB maps, for each of the four methods. In green, we have the detections outside the confidence masks; in blue, detections outside the inpainting mask but inside the common confidence mask; and in salmon, detections within the inpainting mask.}
	\label{f:ps}
\end{figure}

We will close this section by studying the point sources detected by the algorithm. It is interesting to know the intensities of the detections: on the filtered ($\beta$) maps, they have a signal of at least $5.4\sigma_\beta$. The signal of these points before filtering, directly on the CMB maps, can be seen in Fig. \ref{f:ps}; there is not a clear cut on the intensity, since the filtering is able to exploit their shape in order to boost the detecting power.

\begin{figure}
	\centering
	\includegraphics[width=\linewidth]{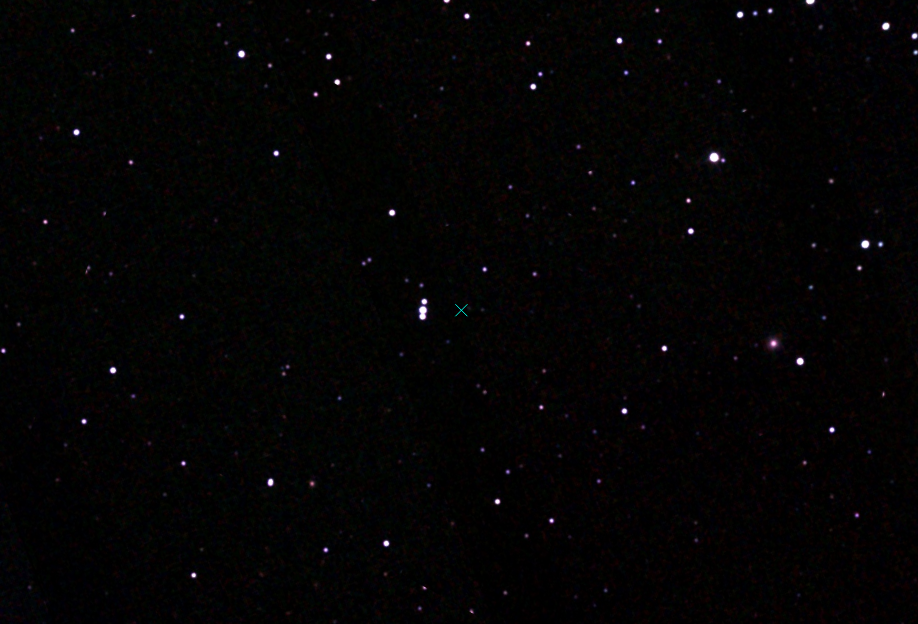}
	\caption{Region of the sky where the point source is detected, in infrared (2MASS). The width of the image is $16'$, while a pixel from Planck is around $1.7'$ wide. The center of the Planck pixel where the maximum is detected is marked with a cyan cross.}
	\label{f:2mass}
\end{figure}

As we mentioned before, there is a point source outside of the confident mask that has been reported by the algorithm both in the SMICA and COMMANDER maps (see Fig. \ref{f:pssmica}; it also corresponds to the green point in Fig. \ref{f:ps}). It can be observed that its shape and intensity are similar to the ones expected from an astrophysical point source. This point is located at galactic longitude $l=103\degree 18'$ and galactic latitude $b=-27\degree 17'$ and, as the majority of the detections, we have to note that this point does not correspond to a source present in the Planck Catalogues of point sources. Identifying this possible source is not straightforward, but looking at this region in an infrared survey such as 2MASS \citep[see][]{skrutskie2006}, we find a bright triplet of point sources within the pixel where the maximum is detected, as it can be seen in \ref{f:2mass}.

\section{Conclusions}

We have implemented the STEM algorithm to extract candidates point sources from a map, controlling the False Discovery Rate. In order to do that, we have written
a code from scratch in Python 3, trying to make our pipeline as flexible as possible, accepting a wide variety of parameters as input.
The code could be used to analyze maps even outside the CMB framework. We have extensively tested the code on CMB simulations, generated using HEALPix and CAMB; we have then the ability to control the proportion of False Detections and the power of the algorithm: we recover most of the sources at intensities higher than 3
to 4 times the standard deviation of the map. Using simulations again, we have tested the effect of noise and masks. We
have concluded that the effect of noise is largely negligible, while masks can introduce some false maxima near the boundary.

We have run the algorithm on a set of foreground cleaned Planck CMB maps, including second and final data release (2015 and 2018, respectively, see i.e., \citealt{planckcollaboration2016e,planckcollaboration2018a}). For the final data release, we have included both inpainted and not inpainted maps (in all cases we focused on temperature anisotropy maps.) We have observed that the inpainting procedure adopted for the 2018 release seems to have produced maps which are much closer to being purely Gaussian, with a number of local maxima largely consistent with theoretical predictions. This was not necessarily the case with some of the earlier releases.

In this paper, we have only focuses on CMB temperature maps. However, these techniques
can be readily applied to any kind of spherical map that we suspect to
be contaminated by point sources. In particular, we plan to apply this algorithm to
polarization maps, both for the E and B modes. Likewise, this algorithm could be eventually applied to frequency maps,
before they are combined to obtain the temperature (or polarization) maps.
In this case, we have the additional problem of diffuse Galactic radiation
contaminating the background, plus all the population of physical point sources. Our aim is also to apply the algorithm to different frequency
bands and then combine the results to improve sensitivity; indeed, in this paper we did not make any use of the spectral information of the CMB
observations.

Finally, we plan to make the code publicly available for the whole scientific community. These and other issues are the objects of ongoing work.

\section*{Acknowledgements}

We acknowledge financial support by ASI grant 2016-24-H.0; the research by DM was also supported by the MIUR Excellence Department Project awarded to the Department of Mathematics, University of Rome Tor Vergata, CUP E83C18000100006. Data analysis was based on observations obtained with Planck (\href{http://www.esa.int/Planck}{http:// www.esa.int/ Planck}), an ESA science mission with instruments and contributions directly funded by ESA Member States, NASA, and Canada. Some of the results in this paper have been derived using the HEALPix \citep{gorski2005} package.


\nocite{adler2009}
\nocite{argueso2011}
\nocite{axelsson2015}
\nocite{bobin2014}
\nocite{cheng2017}
\nocite{delabrouille2013}
\nocite{durastanti2013}
\nocite{lan2009}
\nocite{mayeli2010}
\nocite{mactavish2006}
\nocite{scodeller2012}
\nocite{starck2010}

\bibliography{biblio}

\end{document}